\documentclass[12pt]{iopart}

\usepackage{amssymb}
\usepackage{amsfonts}
\usepackage{bm}
\usepackage{epsfig}
\usepackage{epsf}                          
\usepackage[]{graphicx}  

\usepackage{color}
\usepackage{cite}

\newcommand {\threejot}[6]{\pmatrix{ #1\!\!&#2\!\!&#3\!\cr #4&#5&#6  \cr}}

\newcommand {\brho} {\boldsymbol{\rho}}

\renewcommand {\d} {{\rm d}}
\renewcommand {\i} {{\rm i}}

\renewcommand {\Re} {{\rm Re}}
\renewcommand {\Im} {{\rm Im}}

\newcommand {\ee} {{\rm e}}

\newcommand {\E}  {{\varepsilon}}
\newcommand {\om}  {{\omega}}
\newcommand {\Om}  {{\Omega}}

\newcommand {\bfa} {{\bi a}}
\newcommand {\bfp} {{\bi p}}
\newcommand {\bfr} {{\bi r}}
\newcommand {\bfE} {{\bi E}}

\newcommand {\dip} {{\rm dip}}

\newcommand {\rmA} {{\rm A}}
\newcommand {\rmC} {{\rm C}}
\newcommand {\rmD} {{\rm D}}
\newcommand {\rmF} {{\rm F}}
\newcommand {\rmN} {{\rm N}}
\newcommand {\rmP} {{\rm P}}

\newcommand {\rmSc} {{\rm Sc}}

\newcommand {\calA} {{\cal A}}
\newcommand {\calB} {{\cal B}}

\newcommand {\calD} {{\cal D}}

\newcommand {\calF} {{\cal F}}

\newcommand {\calS} {{\cal S}}

\newcommand {\calV} {{\cal V}}

\newcommand {\walpha} {{\widetilde{\alpha}}}
\newcommand {\wcalA} {\widetilde{\calA}}
\newcommand {\wN} {\widetilde{N}}

\begin{document}
\jl{2}

\title{Vacancy decay in endohedral atoms: the role of 
         non-central position of the atom}

\author{A V Korol$^{1,2}$ and A V Solov'yov$^1$}

\address{$^1$ Frankfurt Institute for Advanced Studies, 
Goethe-Universit\"at, 
Ruth-Moufang-Str. 1, 60438 Frankfurt am Main, Germany}
%%%%%%
\address{$^2$ Department of Physics, 
St.Petersburg State Maritime Technical University, 
Leninskii prospect 101, St.~Petersburg 198262, Russia}

%%%%%%
\eads{korol@fias.uni-frankfurt.de, solovyov@fias.uni-frankfurt.de}

%%%%% Date
\date{today}

%%%%%%%%%%%%%%%%%%%%%%%%%%%%%%%%%%%%%%%%%%%%%%%%%%
\begin{abstract}
We demonstrate that the Auger decay rate in an 
endohedral atom is very sensitive to the atom's location in the fullerene cage.
Two additional decay channels appear in an endohedral system:
(a) the channel due to the change in the electric field at the atom caused by 
dynamic polarization 
of the fullerene electron shell by the Coulomb field of the vacancy,
(b) the channel within which the released energy is transferred to the 
fullerene electron via the Coulomb interaction.
The relative magnitudes of the correction terms are dependent not only 
on the position of the doped atom but also on the transition energy $\om$.
Additional enhancement of the  decay rate appears for 
transitions whose energies are in the vicinity of the fullerene 
surface plasmons energies of high multipolarity.  
It is demonstrated that in many cases the additional channels
can dominate over the direct Auger decay resulting in pronounced broadening 
of the atomic emission lines.
The case study, carried out for Sc$^{2+}$@C$_{80}^{6-}$,
shows that narrow autoionizing resonances in an isolated Sc$^{2+}$ within the range 
$\om = 30\dots 45$ eV
are dramatically broadened if the ion is located strongly off-the-center.
Using the developed model we carry out quantitative analysis of the photoionization
spectrum for the endohedral complex Sc$_3$N@C$_{80}$ and demonstrate that
the additional channels are partly responsible for the strong modification of the
photoionization spectrum profile detected experimentally in \cite{MuellerEtAl2007}.

\end{abstract}

%%%%%%%%%%%%%%%%%%%%%%%%%%%


% 32.70.Jz  Line shapes, widths, and shifts
% 32.80.Fb  Photoionization of atoms and ions 
% 32.80.Zb  Autoionization
% 33.80.Eh  Autoionization, photoionization, and photodetachment
% 36.40.-c  Atomic and molecular clusters 
% 33.70.Jg  Line and band widths, shapes, and shifts

\submitto{\JPB}
%\maketitle
%---------------------------------------------------------------------------
 
%%%%%%%%%%%%%%%%%%
%%%%% Introduction 
%%%%%%%%%%%%%%%%%%%%%%%%%%%%%%%%%%%%%%%%%%%%%%%%
\section{Introduction
\label{Introduction}} 

In this paper, we demonstrate that the rate of atomic non-radiative decay 
in an endohedral system $\rmA@\rmC_{\rmN}$ strongly depends on the location
of the atom inside the fullerene cage.
In particular, the width of the Auger decay can be varied by orders of 
magnitudes by increasing the atomic displacement $a$ from the cage center. 
As $a$ increases the most pronounced enhancement of the width occurs
for the transitions whose energy lies in the vicinity of the fullerene 
surface plasmon of high multipolarity.  

To analyze the influence of the fullerene electron shell on the atomic 
decay we consider two physical mechanisms.
The first one accounts for the change in the electric field at the atom
due to the dynamic polarization of the shell by the  
Coulomb field of the transferring atomic electron \cite{AmusiaBaltenkov2006}.
This scheme is similar to the one which leads to the modification
of  radiative and Auger decays in an isolated atom  due to the intrashell 
many-electron correlations \cite{AmusiaEtAl1972}. 
Within the second mechanism the energy, released in the atomic transition,
is transferred via the Coulomb interaction to the fullerene electron
which becomes ionized.
In  \cite{AverbukhCederbaum2006} this mechanism was called interatomic 
Coulombic decay, thus pointing out its close relationship with the 
interatomic decay process in various molecular clusters 
\cite{CederbaumEtAl1997,SantraEtAl2000,Averbukh2010}.

The influence of the fullerene electron shell on the processes 
of radiative and non-radiative decays of a vacancy in the atom was 
analyzed for spherically symmetric endohedral systems $\rmA@\rmC_{60}$, 
in which the atom $\rmA$ is placed at the center of the spherical cage of
the $\rmC_{60}$ fullerene \cite{AmusiaBaltenkov2006,AverbukhCederbaum2006}.
It was noted that the interatomic Coulombic decay leads to a 
dramatic increase of the atomic decay rate, especially in the case when a 
non-radiative transition in the isolated atom is energetically forbidden.
The estimates presented in the cited papers show that the interatomic decay 
mechanism in $\rmA@\rmC_{60}$ can increase the radiative decay rate by 
a factor $10^5-10^6$.
On the other hand, the impact of the dynamic polarization of the fullerene
on the atomic Auger width was found to be negligibly small
for the at-the-center location of the atom
\cite{AmusiaBaltenkov2006}.

Another prediction made in \cite{AmusiaBaltenkov2006} concerns the modification
of the atomic Auger width due to the interference between a direct wave of 
the electron emitted in the Auger process and its waves scattered from the 
cage.
This effect is identic to the one in the photoionization process of an 
endohedral atom, where the interference of the direct and 
scattered photoelectron waves can lead to the so-called
'confinement resonances' \cite{ConneradeDolmatovManson1999}   
in the photoionization cross section.
In recent years a number of theoretical predictions have been made 
on the properties of the confinement resonances in various  spherical 
endohedral systems (see, e.g., the review \cite{Dolmatov2009}).
However, so far these predictions have not been supported experimentally.
The explanation of the discrepancy between the  predictions and 
experimental results was presented recently \cite{KorolSolov_ConfRes}.
It was shown that the structure of confinement resonances 
in the photoionization cross section of an endohedral atom 
is very sensitive to the mean displacement
$\langle a \rangle $ of the atom from the cage center.
The resonances are strongly suppressed if $2\langle a \rangle$
exceeds the photoelectron half-wavelength. 
Decrease in the amplitude of the confinement resonances with
the increase of the doped atom displacement from the cage center 
was also noted in \cite{BaltenkovBeckerEtAl2010}.
Taking into account similar nature of the confinement resonances in the 
photoionization and Auger decay processes we can state that
the interference effects will not affect the Auger widths if the 
emitted electron wavelength  satisfies the criterion formulated above.

The experimental data on the photoionization of A@C$_N$ 
are sparse due to 
the difficulty to produce sufficient amounts of purified 
endohedrals for the gas phase experiments 
\cite{MuellerEtAl2007,MuellerEtAl2008}.  
The cross sections in the region of giant
atomic resonances were measured for 
Dy@C$_{82}$ \cite{Mitsuke_EtAl2005_Dy@C82},
Ce@C$_{82}$ \cite{Mitsuke_EtAl2005},
Pr@C$_{82}$ \cite{KatayanagiEaAl_2008}
and 
Ce@C$_{82}^{+}$ \cite{MuellerEtAl2007,MuellerEtAl2008,MuellerEtAl_Isacc2009}.

Apart from these the measured cross sections for 
photoionization of the endofullerene Sc$_3$N@C$_{80}^{+}$ in the photon
energy range $30\dots 45$ eV were reported in \cite{MuellerEtAl2007}.
These data are of the prime interest in connection with the topic of the
present paper.
Indeed, as it was shown theoretically 
\cite{AltunManson_1999,SossahZhouManson2008}
and  also confirmed experimentally \cite{SchippersEtAl_2003},
the cross section of photoionization of an isolated ion Sc$^{2+}$
in the indicated energy range is dominated by a set of narrow 
resonances due to the excitations of the 3p 
electrons to intermediate autoionizing states. 
Hence, taking into account that each scandium atom in 
$\rmSc_3\rmN @\rmC_{80}$ carries a positive charge
$q\approx +2 e$ \cite{AlvarezEtAl_2002}, 
it is natural to expect the manifestation of the autoionizing
 resonances in the photoionization of the endofullerene.
However, such distinct structure has not been seen in the experiment
\cite{MuellerEtAl2007}. 
Instead, the presence of the endohedral molecule  $\rmSc_3\rmN$ 
resulted in a single wide peak 
(of  6 eV  full width at half maximum) 
in the photoionization curve of $\rmSc_3\rmN @\rmC_{80}$.   
In \cite{MuellerEtAl2007} it was mentioned that such a dramatic
modification can be due to the presence of the fullerene cage 
which could cause significant broadening of any atomic resonance 
features in photoionization.

The intriguing experimental results of \cite{MuellerEtAl2007} 
has stimulated our present study on the dependence of the atomic non-radiative decay
processes on the doped atom displacement from the cage center.
For doing this we developed a formalism which allows one to 
determine the corrections to the atomic Auger width due to the 
dynamics of the fullerene electrons.
Within the framework of the formalism, which is presented in
\sref{Formalism}, one accounts for the off-the-center position of 
the doped atom and expresses the corrections in terms of the 
fullerene multipole dynamic polarizabilities of the second kind.
On the basis of the developed formalism in \sref{Numerics} 
we carry out quantitative analysis of the dependence of the autoionizing 
width on the displacement $a$.
This is done for the Sc$^{2+}$ ion encaged in the fullerene C$_{80}$.
The obtained data are used further to calculate the photoionization
spectrum for the endohedral complex Sc$_3${\rm N@C}$_{80}^{+}$ 
in the photon energy range $30\dots 45$ eV.
In \ref{MultipolPolar} we discuss different types of polarizabilities 
of the hollow objects and present the model description of the dynamic 
multipole polarizabilities of a fullerene.

The atomic system of units is used through the paper.

%%%%%%%%%%%%%%%%%%
%%%%% Formalism
%%%%%%%%%%%%%%%%%%%%%%%%%%%%%%%%%%%%%%%%%%%%%%%%
\section{The formalism 
\label{Formalism}}

%%%%%%%%%%%%
\subsection{Auger width in an isolated atom
\label{Auger_isolated_atom}} 

For the sake of clarity and consistency of the notations and terminology 
adopted below in the paper, let us outline basic formalism related
to the calculation of the Auger width in an isolated atom.

\begin{figure}[h]
\centering
\parbox{20cm}{
\hspace*{2cm}
\includegraphics[clip,scale=0.40]{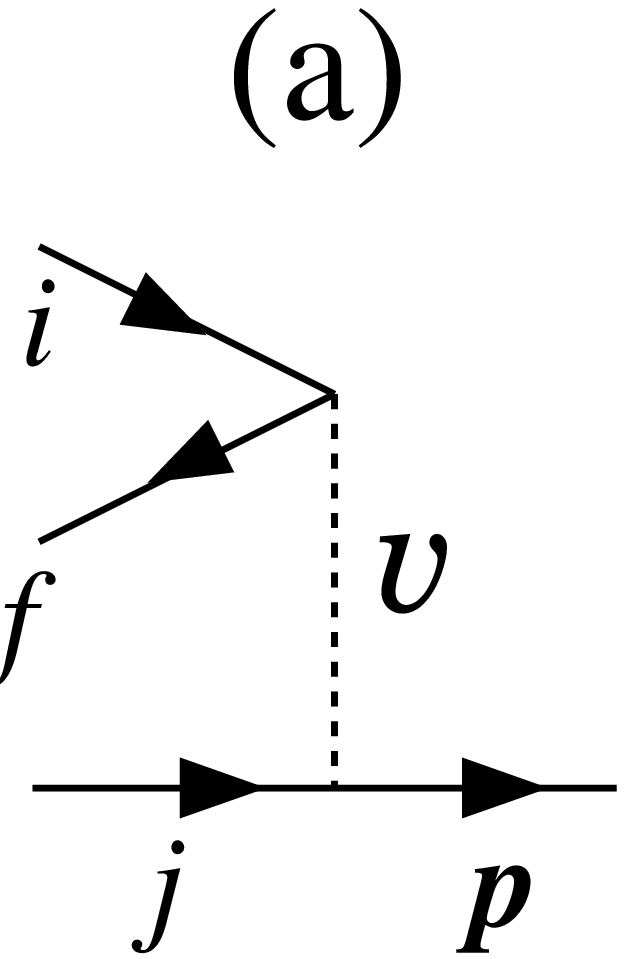}%{diagram1a.eps}
\hspace*{1cm}
\includegraphics[clip,scale=0.40]{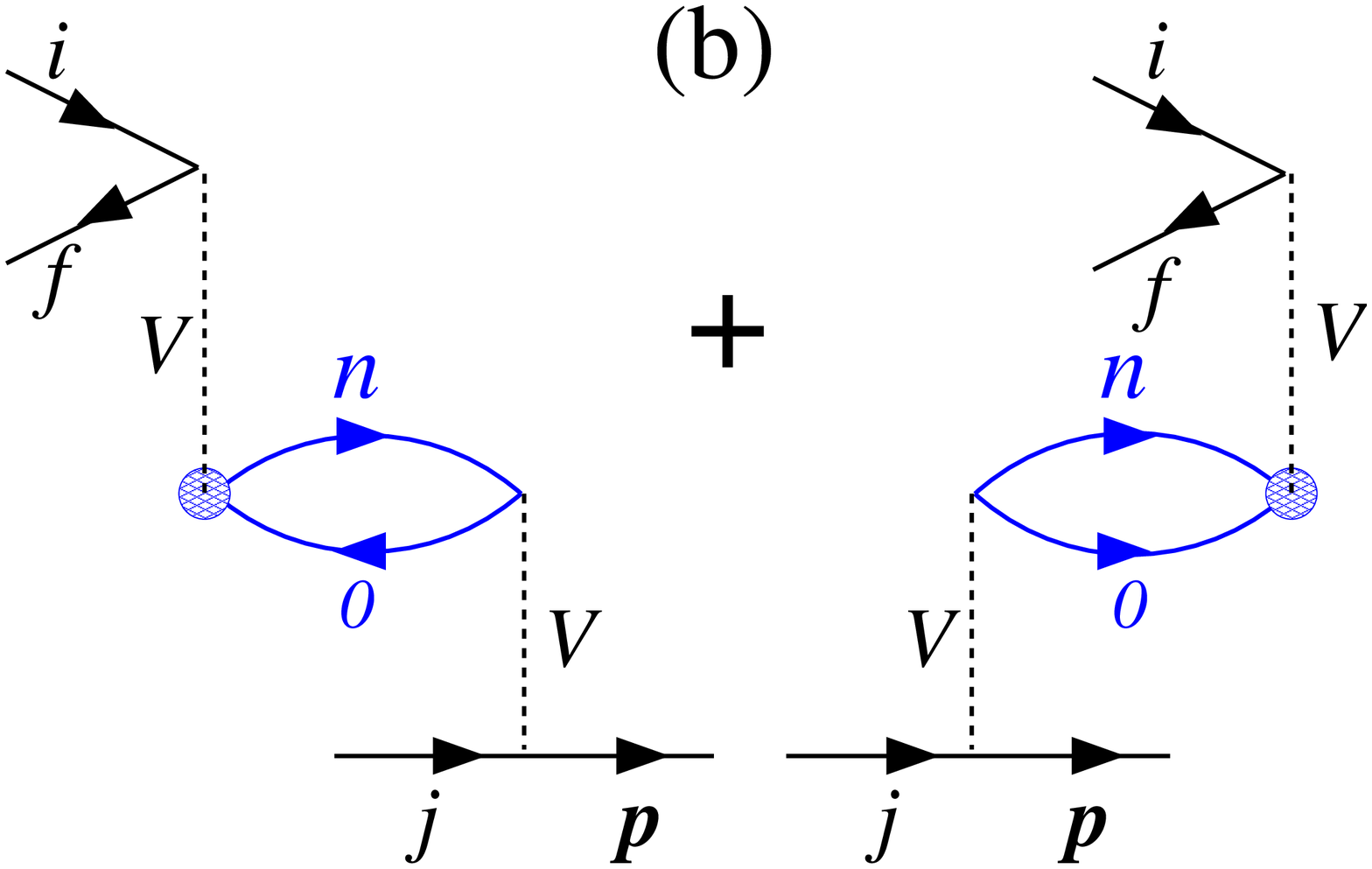}%{diagram2a.eps}
}
\caption{
{\bf (a)} Diagrammatical representation of the Auger 
decay in an isolated atom. 
Both transitions $i \to f$ and $j \to \bfp$ occur with atomic electrons.
Notation $v$ stands for the (effective) Coulomb interaction.
\\
{\bf (b)} Additional channel in the endohedral atom:
the decay 
occurs via the virtual polarization of the fullerene 
(the  particle-hole excitation $0\to n$).
$V$ stands for the  atom-fullerene Coulomb interaction, the filled 
circle denotes 
the effective vertex of the particle-hole excitation in the fullerene.
}
\label{diagram1.fig}
\end{figure}

The Auger decay of the vacancy in an isolated atom
can be illustrated by the diagram presented in figure \ref{diagram1.fig}(a).
The inner vacancy $f$ is filled by the electron from the outer state
$i$. 
Due to the interelectron Coulomb interaction $v=1/|\bfr-\bfr^{\prime}|$ 
the released energy $\E_i-\E_f$
is transferred to another electron from state $j$ which becomes ionized
($\bfp$ stands for the asymptotic
momentum of the outgoing electron).
The energy conservation implies $\E_i+\E_j = \E_f + \E$, where
$\E=p^2/2$. 

%%%%%%%%%%%%%%%%%%%%%%%%%% Figures fig1a.eps and fig1b.eps %%%%%%
%%%%%%%%%%%%%%%%%%%%%%%%%%%%%%%%%%%%%%%%%%%%%%%%%%%%%

The amplitude $A_{ij\to f\bfp}^{\rmA}$ of the process from 
figure \ref{diagram1.fig}(a) is proportional to the two-electron Coulomb
matrix element $\langle f \bfp| v | j i \rangle$.
Presenting the wavefunctions of single-electron states $i,j,f$ as 
products of angular and radial parts, and using the standard multipole 
expansions of $v$ and 
of the outgoing electron wavefunction, one derives the analytical
expression for  $A_{ij\to f\bfp}^{\rmA}$ (the subscript '$\rmA$' stresses
that the quantity refers to the isolated atom). 
In what follows, for the sake of concreteness, we focus on the 
{\em dipole} transitions $i \to f$ and $j \to \bfp$.
As it is discussed further in the paper, the presence of the fullerene
cage, which radius noticeably exceeds that of the endohedral atom,
 most strongly affects the dipole Auger transitions.  
In the dipole approximation the amplitude can be written as follows:
%%%%%%%%%%%%%%%
\begin{eqnarray}
A_{ij\to f\bfp}^{\rm dip,\, \rmA}
&=
-\delta_{s_fs_i}\delta_{s s_j}\,
\sqrt{8\pi^3  \over p}
\sum_{l=l_j\pm 1}
\i^{-l}\,\ee^{\i \delta_{\nu}}\,
\calV^{(1)}_{\nu_f\nu;\nu_i\nu_j}\,
\nonumber\\
&\quad
\times
(-1)^{m_f+m_j} 
\sum_{m\mu} 
\threejot{\!  l_f}{\!    1}{\!  l_i}
         {\! -m_f}{\!  \mu}{\!  m_i}
\! 
\threejot{\!  l_j}{\!    1}{\!  l}
         {\! -m_j}{\!  \mu}{\!  m}
\! 
Y_{lm}(\bfp)\,,
%T_{m_fm_im_j}^{l_fl_il_jl}(\bfp)\, .
\label{Auger_dipole.1}
\end{eqnarray}
where $\threejot{a}{b}{c}{\alpha}{\beta}{\gamma}$ stands for the
$3j$-symbol.
The non-indexed quantities on the right-hand side
refer to the outgoing electron, whereas the 
indexed ones (the subscripts $a=i,j,k$) - to the states $i,j,k$. 
The factors $\delta_{s_fs_i}\delta_{s s_j}$ take into account that 
within the non-relativistic framework the spin projections $s$ are not
changed in the transitions $i \to f$ and $j \to \bfp$.
The notation $\nu_a \equiv (n_a,l_a)$ stands for the set of principal
and orbital quantum numbers, $\nu \equiv (\E,l)$, and
$\delta_{\nu}$ are the scattering phaseshifts.
The quantity $\calV^{(1)}_{\nu_f\nu;\nu_i\nu_j}$ denotes the dipole radial
Coulomb matrix element:
\begin{eqnarray}
\fl
\calV^{(1)}_{\nu_f\nu;\nu_i\nu_j}
&=
\left.
\Pi_{l_fl_ill_j}\!
\threejot{\! l_f}{\! 1}{\! l_i}
         {\!   0}{\! 0}      {\!   0}
\!\!
\threejot{\! l_j}{\! 1}{\! l}
         {\!   0}{\! 0}{\! 0}
\!\!
\int\limits_{0}^{\infty}
\!\!
\int\limits_{0}^{\infty} 
\d r \,
 \d r^{\prime}\, 
P_{\nu_f}(r) P_{\nu}(r^{\prime})\,
{r_{<}^{\lambda} \over r_{>}^{\lambda+1}}\,
P_{\nu_i}(r)P_{\nu_j}(r^{\prime})
\right|_{\lambda=1}
\!\!,
\label{Auger.6}
\end{eqnarray}
where $\Pi_{l_al_bl_c\dots}=\sqrt{(2l_a+1)(2l_b+1)(2l_c+1)\dots}$.

Equation (\ref{Auger_dipole.1}) has been 
derived in a single-electron approximation.
To account for the electron correlations  one substitutes 
$\calV^{(1)}_{\nu_f\nu;\nu_i\nu_j}$ with the corresponding
matrix element of the effective interaction 
(see, e.g., \cite{Amusia_PI_Book}).

The width, which defines  
the probability (per unit time) of the Auger transition,
one calculates starting from the following general formula:
\begin{eqnarray}
\Gamma_{ij\to f\E}
&=
{p \over 8\pi^2 \Pi_{l_i}^2}
\sum_{m_i m_f m_j}
\sum_{s_i s_f \atop s_j s}
\int \d \Om_{\bfp} 
\left|A_{ij\to f\bfp}
\right|^2
=
\sum_{l} \Gamma_{\nu_f\nu;\nu_i\nu_j},
\label{width1.1}
\end{eqnarray}
where $\Gamma_{\nu_f\nu;\nu_i\nu_j}$ stands for the partial width of the
transition $\nu_i\nu_j\to \nu_f\nu$. 

Using (\ref{Auger_dipole.1}) in (\ref{width1.1}), one derives the following 
expression for the  partial width of the 
dipole Auger transition in the isolated atom:
\begin{eqnarray}
\Gamma_{\nu_f\nu; \nu_i \nu_j}^{\rm dip,\, A}
&=
{4\pi \over 3\Pi_{l_i}^2}\,
\Big|\calV^{(1)}_{\nu_f\nu;\nu_i\nu_j}\Bigr|^2 \,.
\label{width2.2}
\end{eqnarray}

%%%%%%%%%%%%
\subsection{Correction to the width of the Auger decay in the endohedral atom
\label{Problem}} 

If the atom is encaged in a fullerene C$_{\rm N}$, the atomic decay process  
can be strongly influenced due to the excitations (real or virtual) of the 
fullerene electrons \cite{AmusiaBaltenkov2006,AverbukhCederbaum2006}.

To start with we mention, that 
the atomic Auger decay can occur via the additional channel, 
diagrammatical representation of which is given in figure \ref{diagram1.fig}(b)
\cite{AmusiaBaltenkov2006}.
Here, the energy $\om_{if} = \E_i-\E_f$ released in the transition $i\to f$ 
is transferred to the electron $j$ not directly, as in the process 
from figure  \ref{diagram1.fig}(a), but via the virtual
excitation of the fullerene. 
The diagrams \ref{diagram1.fig}(b) constitute the correction term, 
$\Delta A_{ij\to f\bfp}^{\rm A@C_N}$, to the amplitude of the 
Auger decay.
Hence, the total amplitude
of the Auger decay in the encapsulated atom reads
\begin{eqnarray}
A_{ij\to f\bfp}^{\rm A@C_N}
=
A_{ij\to f\bfp}^{\rmA}
+
\Delta A_{ij\to f\bfp}^{\rm A@C_N}\,.
\label{total_Auger.1}
\end{eqnarray}

Physically, the virtual excitation implies that the Coulomb interaction
between two atomic electrons, $i$ and $j$, is modified due to the 
polarization of the fullerene shell.
The polarization is dynamic, i.e. it depends on the transition energy 
$\om_{if}$, 
so that $\Delta A_{ij\to f\bfp}^{\rm A@C_N}$ is proportional to the dynamic 
susceptibility of the fullerene.
Thus, one can expect, that in those $\om_{if}$-regions 
where the modulus of susceptibility is large enough,  
the additional channel can modify noticeably  the width of the
transition $ij\to f\bfp$.

%%%%%%%%%%%%%%%%%%%%%%%%%% Figure diagram2.eps %%%%%%
%\begin{figure}[h]
%\centering
%\parbox{10cm}{
%\includegraphics[clip,scale=0.40]{diagram2.eps}
%}
%\caption{
%Additional channel of the atomic Auger decay $ij \to f\bfp$
%in the endohedral atom:
%the decay 
%occurs via the virtual polarization of the fullerene 
%(the  particle-hole excitation $0\to n$).
%$V$ stands for the  atom-fullerene Coulomb interaction, the filled 
%circle denotes 
%the effective vertex of the particle-hole excitation in the fullerene.
%}
%\label{diagram2.fig}
%\end{figure}
%%%%%%%%%%%%%%%%%%%%%%%%%%%%%%%%%%%%%%%%%%%%%%%%%%%%%

In \cite{AmusiaBaltenkov2006} the additional channel
of the dipole Auger decay was analyzed for an atom placed 
{\it at the center} of a spherically-symmetric fullerene.  
Assuming the atomic radius $R_a$ to be much smaller than the fullerene 
radius $R$, the authors expressed the amplitude 
$\Delta A_{ij\to f\bfp}^{\rm \dip, A@C_N}$ in terms of dynamic 
dipole polarizability of the fullerene. 
Then, carrying out the order-of-magnitude estimate they found that
$\Bigl|\Delta A_{ij\to f\bfp}^{\rm dip, A@C_N}/
A_{ij\to f\bfp}^{\dip, \rmA}\Bigr|
\sim \Bigl(R_a/R\Bigr)^3 \ll 1$, i.e., the correction 
is much smaller that the Auger amplitude in the isolated atom in the whole
range of $\om$.   

Below in the paper we demonstrate, that the relative magnitude of the 
correction term is governed not only by $\om$ (this has been already
noted in \cite{AmusiaBaltenkov2006,AverbukhCederbaum2006}) but 
also is strongly  dependent on the position of the atom inside the cage.
The parameter, which defines this dependence, is the ratio $a/R$, 
where $a$ stands for the displacement of the
atom from the cage center.  
Below in this section we demonstrate, that the correction term
as a function of both $\om_{if}$ and $a/R$ varies by orders of magnitude.
This can result in the dominance of the additional channel over the
direct Auger decay (see the case study presented in section 
\ref{Numerics_Sc2+}).

\begin{figure}[h]
\centering
\includegraphics[clip,scale=0.3]{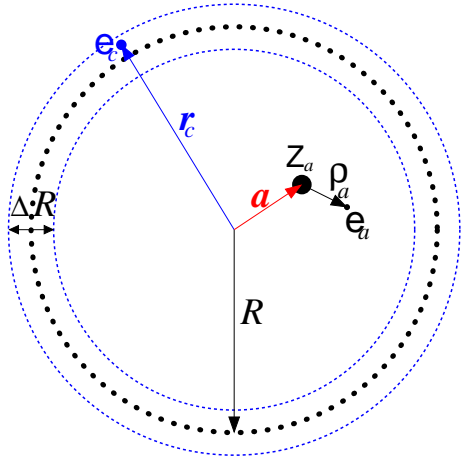}%{atom_cage.eps}
\caption{
Illustrative picture of the atom located inside the 
spherical fullerene shell.
The shell average radius and width are notated as $R$ and $\Delta R$,
respectively. 
The atomic nucleus (the charge $Z_a$) is displaced by vector $\bfa$ 
from the center.
Position vectors of the fullerene, $e_c$, and the atomic, $e_a$, 
electrons are notated as 
$\bfr$ and $\brho$.
}
\label{atom-cage.fig}
\end{figure}

Let us outline the derivation of the correction term 
$\Delta A_{ij\to f\bfp}^{\rm dip,\, A@C_N}$ for the dipole Auger 
transition occurring in the atom (nucleus charge $Z_a$, average radius $R_a$),
located inside spherically-symmetric fullerene 
(average radius of the cage $R$, average width of the shell $\Delta R$) 
and displaced by $\bfa$ from the cage center,
see figure \ref{atom-cage.fig}.

%%%%%%%%%%%%%%%%%%%%%%%%%% Figure atom-cage
%%%%%%%%%%%%%%%%%%%%%%%%%%%%%%%%%%%%%%%%%%%%%%

Two diagrams in figure \ref{diagram1.fig}(b) correspond to the 
following analytic expression:
\begin{eqnarray}
\fl
\Delta A_{ij\to f\bfp}^{\rm A@C_N}(a)
&=
\delta_{s_fs_i}\delta_{s s_j}
\sum_{n}
%\sum_{\nu_0 \nu_n}\sum_{m_0 m_n}
\left[
{\langle f; n| V | 0; i \rangle  \langle \bfp; 0| V | n; j \rangle 
\over \om_{n0} - \om_{if}}
+
{\langle f; 0| V | n; i \rangle  \langle \bfp; n| V | 0; j \rangle 
\over \om_{n0} + \om_{if}
}
\right] .
\label{d_Auger.4}
\end{eqnarray}
The sum is carried out over the complete spectrum of the 
fullerene excited states 
(the integration over the continuous spectrum is implied),
$\om_{n0}=\E_n-\E_0$ is the energy of the virtual transition $0\to n$.

The quantity $V=\sum_{a,c}1/ |\bfr_c - \bfa - \brho_a|$ stands for the Coulomb
interaction between the atomic and fullerene electrons,
see figure \ref{atom-cage.fig}. 
Assuming the fullerene electrons to be located farther
from the center than atomic electrons, i.e. $r > |\bfa + \brho|$,  
and choosing the $z$-axis along $\bfa$
one expresses $V$ in terms of multipole series 
\cite{VMX,IliaSolovyov_EtAl2007}: 
\begin{eqnarray}
\fl
V 
\approx 
\sum_{l=0}^{\infty}\!
\sum_{l_1,l_2=0\atop l_1+l_2=l}^l \!
a^{l_1} \,
{\sqrt{(2l+1)!} \over \sqrt{(2l_1)!(2l_2)!} }
\sum_{m m_2}
(-1)^{m}
\!
\threejot{ l_1}{ l_2}{  l}
         {   0}{ m_2}{ -m}
C_{lm}^{*}(\{\bfr_c\})\, 
Q_{l_2m_2}(\{\brho_a\})\,.
\label{atom-cage.4}
\end{eqnarray}
Here, the quantity $Q_{lm}$, defined as
\begin{eqnarray}
Q_{lm}(\{\brho_a\})
= 
\sqrt{4\pi\over 2l+1}\,
\sum_a\rho_a^{l}\, Y_{lm}(\brho_a)\,,
\label{atom-cage.5}
\end{eqnarray}
is commonly termed as the operator of $2^l$-pole electric moment of 
the atom (the sum is carried out over atomic electrons). 
The quantity 
\begin{eqnarray}
C_{lm}(\{\bfr_c\})
= 
\sqrt{4\pi\over 2l+1}\,
\sum_c {Y_{lm}(\bfr_c) \over r_c^{l+1}}\,,
\label{atom-cage.6}
\end{eqnarray}
also is related to the multipole expansion of the electrostatic 
field created by electron charge  distributed in the fullerene shell
(the sum is carried out over the fullerene electrons).
To distinguish $C_{lm}$ from $Q_{lm}$ we call the former as 
the operator of "interior $2^l$-pole moment" of the fullerene
(see \ref{MultipolPolar_1} for the details).

The argument $a$ on the left-hand side of (\ref{d_Auger.4}) is added to
stress that the amplitude is dependent on the displacement from the
center.

Inserting (\ref{atom-cage.4}) into (\ref{d_Auger.4}), one carries out the
intermediate algebra and arrives at the following expression 
for the amplitude of the {\it dipole}  $i \to f$ and  $j \to \bfp$
transitions: 
\begin{eqnarray}
\fl
\Delta A_{ij\to f\bfp}^{\dip,\, \rm A@C_N}(a)
&=
\delta_{s_fs_i}\delta_{s s_j}(-1)^{m_f+m_j}
\sqrt{8\pi^3 \over p}
d_{\nu_f\nu_i}^{\rm (A)}
\sum_{l=l_j\pm 1} 
\i^{-l}\,\ee^{\i \delta_{\nu}}\,
d_{\nu\nu_j}^{\rm (A)} 
\nonumber\\
\fl
&
\times 
\sum_{L=1}^{\infty}
a^{2L-2}
{(2L+1)! \over 2(2L-2)! }\,
\walpha_{L}^{\rm (C_N)}(\om_{if})
\sum_{m M \atop  \mu_1\mu_2}
\threejot{\! l_f}{\!     1}{\! l_i}
         {\!-m_f}{\! \mu_1}{\! m_i}
\nonumber\\
\fl
&
\times
\threejot{\! L-1}{\!     1}{\!  L}
         {\!   0}{\! \mu_1}{\! -M}
\!
\threejot{\! L-1}{\!      1}{\!  L}
         {\!   0}{\! -\mu_2}{\!  M}
\!
\threejot{\!  l_j}{\!    1}{\! l}
         {\! -m_j}{\!\mu_2}{\! m}
Y_{lm}(\bfp)\,.
\label{d_Auger_non_central4.1}
\end{eqnarray}
Here 
$d_{\nu_f\nu_i}^{\rm (A)}$ and $d_{\nu\nu_j}^{\rm (A)}$ stand for the radial 
matrix elements of the atomic dipole moment,
$\walpha_{L}^{\rm (C_N)}(\om_{if})$ is the fullerene's dynamic  
$2^L$-pole polarizability of the second kind.
This type of polarizability appears when one is interested in the 
modification of the electric field inside the fullerene due to 
its polarization under the action of the $2^L$-pole external field whose
source is also located in the fullerene interior.
This is in contrast to the "conventional" polarizability
$\alpha_{L}^{\rm (C_N)}(\om_{if})$ which is responsible for the same effect 
but in the case where both the source and the observation point are located 
outside the fullerene
(see \ref{MultipolPolar} for more details).

Summing up (\ref{Auger_dipole.1}) and (\ref{d_Auger_non_central4.1}) 
one constructs the total amplitude 
$A_{ij\to f\bfp}^{\dip,\, \rm A@C_N}$ of the dipole
Auger decay. 
Then, using $A_{ij\to f\bfp}^{\dip,\, \rm A@C_N}$ in (\ref{width1.1}), 
one derives the following set of expressions, which defines the 
partial width of the transition
$\nu_i\nu_j\to \nu_f \nu$ in the encaged atom:
\begin{eqnarray}
\Gamma_{\nu_f\nu; \nu_i\nu_j}^{\rm dip,\, A@C_N}(a)
= 
\Gamma_{\nu_f\nu; \nu_i\nu_j}^{\rm dip,\, A}\,  
F_{\nu_f\nu; \nu_i\nu_j}(a)\,.
\label{d_Auger_non_central5.11}
\end{eqnarray}
The factor $F_{\nu_f\nu; \nu_i\nu_j}(a)$, which depends on the atom's 
displacement, takes into account the width 
modification due to the presence of the fullerene (for an isolated atom
$F_{\nu_f\nu; \nu_i\nu_j}(a)\equiv 1$).
It can be written in the form
\begin{eqnarray}
\fl
F_{\nu_f\nu; \nu_i\nu_j}(a)
=
1 - 2\kappa_{\nu_f\nu;\nu_i\nu_j}\,\Re\, \Sigma_1(\om_{if};a)
+
\kappa_{\nu_f\nu; \nu_i\nu_j}^2
\left(
\Bigl| \Sigma_1(\om_{if};a)\Bigr|^2
+
{\Bigl| \Sigma_2(\om_{if};a)\Bigr|^2 \over 2}
\right).
\label{d_Auger_non_central5.12}
\end{eqnarray}
The factor $\kappa_{\nu_f\nu;\nu_i\nu_j}$ is constructed from
the atomic characteristics only: 
\begin{eqnarray}
\kappa_{\nu_f\nu; \nu_i\nu_j}
=
\left(
{\Gamma_{\nu_f\nu_i}^{\gamma, {\rm A}}\over 
 \Gamma_{\nu_f\nu;\nu_i\nu_j}^{\rm dip, A}}
\,
\sigma_{\nu_j}(\om_{if})
\right)^{1/2},
\label{d_Auger_non_central5.13a}
\end{eqnarray}
 where 
$\Gamma_{\nu_f\nu_i}^{\gamma, {\rm A}}$ is the partial radiative
width of the transition $\nu_i \to \nu_f$,
%$\sigma_{\nu\,\nu_j}^{\rm (A)}(\om_{if})$ 
$\sigma_{\nu_j}(\om_{if})$ 
is the  partial 
cross section of photoionization of the subshell $\nu_j$, 
and $ \Gamma_{\nu_f\nu;\nu_i\nu_j}^{\rm dip, A}$ is given by
(\ref {width2.2}). 

The quantities $ \Sigma_{1,2}(\om_{if};a)$ are dependent on the 
displacement $a$ and on 
the fullerene's dynamic multipole polarizabilities of the second kind:
\begin{eqnarray}
\left\{
\begin{array}{c}
\Sigma_1(\om_{if};a)
\\
\Sigma_2(\om_{if};a)
\end{array}
\right\}
=
\sqrt{3 \over 8 \pi}  
\,
{c^2 \over 3\om_{if}^2}
\sum_{l=1}^{\infty}
\left\{
\begin{array}{c}
l (2l+1)
\\
l(l-1)
\end{array}
\right\}
a^{2l-2} \,
\walpha_{l}^{\rm (C_N)}(\om_{if})\,,
\label{d_Auger_non_central5.13b}
 \end{eqnarray}
where $c\approx 137$ is the light velocity in the atomic units.
As a function of the fullerene cage radius $R$ the polarizability 
$\walpha_{l}^{\rm (C_N)}(\om_{if})$ scales as $\sim R^{2l-4}$
(see (\ref{MultipolPolar_2.02}) and (\ref{ModifiedPRA.6})).
Therefore, the expansion parameter in the series 
(\ref{d_Auger_non_central5.13b}) is $(a/R)^2$.

In the limit $a=0$ (atom at-the-center) only dipole terms with $l=1$ contribute
to the sums (\ref{d_Auger_non_central5.13b}).
Hence, one finds  $\Sigma_2(\om_{if};0)=0$ and
\begin{eqnarray}
\Sigma_1(\om_{if};0)
=
\sqrt{3 \over 8 \pi}  
\,
{c^2 \over \om_{if}^2}
\walpha_{1}^{\rm (C_N)}(\om_{if})
\approx
\sqrt{3 \over 8 \pi}  
\,
{c^2 \over \om_{if}^2}
{\alpha^{\rm (C_N)}_1(\om_{if}) \over R^6},
\label{d_Auger_non_central5.14a}
\end{eqnarray}
where $\alpha^{\rm (C_N)}(\om_{if})$ is the dynamic dipole polarizability of 
the fullerene (the transformation from $\walpha_{1}^{\rm (C_N)}(\om_{if})$
to $\alpha^{\rm (C_N)}_1(\om_{if})$, which leads to the approximate 
equality, is explained in \ref{Appendix_B}). 
As a result, the factor (\ref{d_Auger_non_central5.12}) reduces to 
\begin{eqnarray}
F_{\nu_f\nu; \nu_i\nu_j}(0)
&\approx
\left| 
1
-
\kappa_{\nu_f\nu; \nu_i\nu_j}\,
\sqrt{3 \over 8 \pi}  
\,
{c^2 \over \om_{if}^2}\,
{\alpha^{\rm (C_N)}(\om_{if}) \over R^6}
\right|^2\,.
\label{d_Auger_non_central5.14b}
\end{eqnarray}

Expression (\ref{d_Auger_non_central5.14b}) was obtained in 
 \cite{AmusiaBaltenkov2006}. 
The authors, estimating the magnitude of the correction term in the 
brackets as $(R_a/R)^3\ll 1$, 
derived $F_{\nu_f\nu; \nu_i\nu_j}(0) \approx 1$.
Thus, it was tacitly concluded that the additional channel
of the atomic Auger decay, figure \ref{diagram1.fig}(b), does not affect 
the decay rate if the atom is placed at the center.

However, as the displacement $a$ from the cage center increases, the
larger contribution to the sums from (\ref{d_Auger_non_central5.13b})
comes from the terms with higher $l$.
This leads to a noticeable increase in magnitude 
of the factor $F_{\nu_f\nu; \nu_i\nu_j}(a)$ not only in the vicinity
of the fullerene giant plasmon resonance but in a much wider range 
of transition energies $\om_{fi}$.
Potentially, the atomic Auger decays with $\om_{fi}$ within the
interval $10\dots 40$ eV can be affected.

\begin{figure}[h]
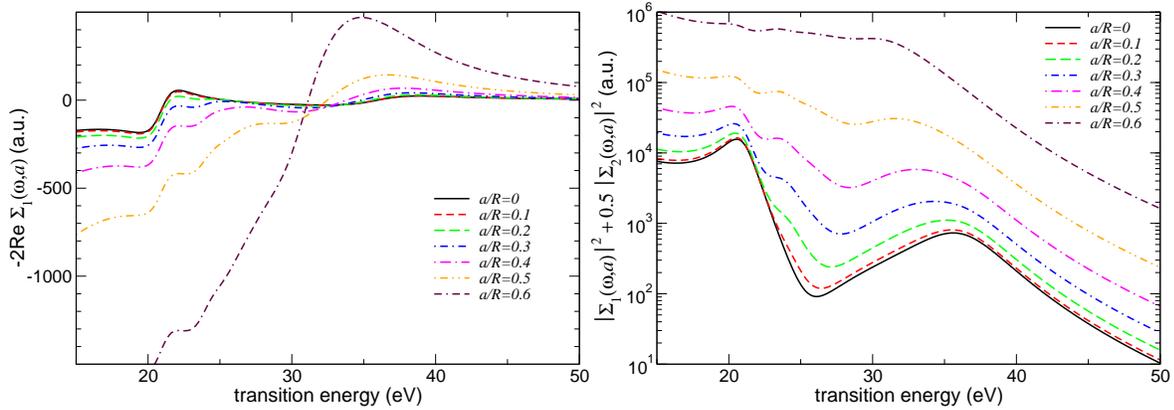

\includegraphics[clip,scale=0.31]{fig3a.eps}%{2ReSigma1_C80_g2_020.eps}
\includegraphics[clip,scale=0.31]{fig3b.eps}%{Sigma1and2_C80_g2_020.eps}
\caption{
Dependences 
$f_1(\om;a) \equiv-2\Re \, \Sigma_1(\om;a)$
(left panel) 
and 
$f_2(\om;a) \equiv\Bigl| \Sigma_1(\om;a)\Bigr|^2
+
\Bigl| \Sigma_2(\om;a)\Bigr|^2/2$
(right panel) 
on the transition energy $\om$ calculated for various $a/R$ 
values as indicated 
(see equations (\ref{d_Auger_non_central5.12}) and 
(\ref{d_Auger_non_central5.13b})).
The data refer to the fullerene C$_{80}^{6-}$
(see \ref{C80}) and explanations in the text.
}
\label{Sigma_C80.fig}
\end{figure}

To illustrate this statement, in figure \ref{Sigma_C80.fig} 
we plot the dependences of $f_1(\om;a) \equiv -2\Re \, \Sigma_1(\om;a)$
and 
$f_2(\om;a) \equiv 
\Bigl|\Sigma_1(\om;a)\Bigr|^2+\Bigl|\Sigma_2(\om;a)\Bigr|^2/2$
(see  (\ref{d_Auger_non_central5.12}))
on the transition energy $\om$ and for several $a/R$ ratios.
The calculations were performed for the fullerene ion C$_{80}^{6-}$.
For the cage radius we used $R=4.15$ \AA\,  \cite{NakaoKutitaFujita_1994},
the width of the shell was fixed at $1.5$ \AA.
The fullerene polarizabilities $\walpha^{\rm (C_N)}_l(\om)$ were calculated 
within the framework of the plasmon resonance approximation 
(see, e.g., \cite{SolovyovConnerade2002,Solovyov2005}) 
and with account for two coupled surface plasmons excited in a fullerene 
(see, e.g., 
\cite{KorolSolovyov2007,LoKorolSolovyov2007,LoKorolSolovyov2009} and 
references therein).
The details of the PRA formalism in describing the dynamic polarizabilities 
of the shell of a finite width are given in \ref{PRA}.
The list of  cage-dependent parameters,  used in the calculations, 
one finds in \ref{C80}.

%%%%%%%%%%%%%%%%%%%%%%%%%% Figure fig3a.eps and fig3b.eps

By itself large absolute values of the functions $f_1(\om;a)$ and/or
$f_2(\om;a)$ in the energy interval $\approx 10\dots 40$ eV   
do not imply that the last two terms on the right-hand-side of 
(\ref{d_Auger_non_central5.12}) will result in a strong modification of the
Auger width.
Indeed, these correction terms are dependent also on the factor 
$\kappa_{\nu_f\nu; \nu_i\nu_j}$, whose magnitude is determined by
atomic characteristics taken at given energy (see equation
(\ref{d_Auger_non_central5.13a})).
Nevertheless, studying the curves in figure \ref{Sigma_C80.fig} 
one  can speculate that, as the displacement increases, the region 
$\om \approx 30 \dots 40$ eV becomes favourable from the viewpoint  
the Auger width modification due to the channel depictured in 
figure \ref{diagram1.fig}(b).
Within the indicated interval not only strong inequality $f_2(\om;a)\gg 1$ 
is valid but also $f_2(\om;a) >0$, which means that the interference
of the two Auger decay pathways, figures \ref{diagram1.fig}(a) and 
\ref{diagram1.fig}(b), is positive. 
This leads to the enhancement of the decay rate. 

For a centrally positioned atom the enhancement is more pronounced
at $\om\approx 20$ eV (see the solid curve in both graphs in figure
\ref{Sigma_C80.fig}.
This energy corresponds to the dipole plasmon resonance associated with the 
symmetric mode of the coupled oscillations of the charges induced at the
fullerene surfaces (see  \ref{PRA}). 
The second surface plasmon peak, which is due to the asymmetric 
mode of the oscillations, is less pronounced in the profile of
the dipole polarizability of the second kind 
$\walpha^{\rm (C_N)}_1(\om)$ (see figure \ref{A_L_C80.fig}).
As the multipolarity $l$ increases, the resonance energies move
towards each other and their intensities gradually equalize.
For $l\gg 1$ the resonances merge at $\om \approx 30$ eV.
This explains the increase of $f_1(\om;a)$ and $f_2(\om;a)$ with $a$:
as the displacement increases, the terms with $l\gg 1$ contribute 
more to the series (\ref{d_Auger_non_central5.13b}).

Therefore, a good candidate for the Auger width modification due to the 
mechanism from figure \ref{diagram1.fig}(b) would be an atom whose Auger
spectrum has pronounced peaks in the range $\om \approx 30 \dots 40$ eV,
and which, being encaged, is located far off-the-center.
The molecule Sc$_3$N@C$_{80}^{+}$ is a good example of this sort.
Indeed, three scandium atoms are symmetrically located far from the center 
being displaced by $a\approx2$ \AA\, 
\cite{StevensonEtAl1999,SlaninaNagase2005}.
In this complex, each scandium atom donates 2 electrons to the fullerene cage.
Hence, the theory developed above in this section can be applied to 
the Auger transitions in the ion Sc$^{2+}$, for which the Auger spectrum 
has been investigated both experimentally \cite{SchippersEtAl_2003}
and theoretically \cite{AltunManson_1999,Kjeldsen2006,SossahZhouManson2008}.
In section \ref{Numerics_Sc2+} we consider the modification of the Auger
widths in  Sc$^{2+}$ due to the presence of the C$_{80}^{6-}$ shell.
The obtained results are applied further in section \ref{Numerics_Sc3N@C80}
to calculate the photoionization spectrum of the Sc$_3$N@C$_{80}^{+}$
molecule. 
The results of our calculations are compared with recent experiment
\cite{MuellerEtAl2007}.
 
%%%%%%%%%%%%
\subsection{Additional channel of the Auger decay in the endohedral atom
\label{additional}} 

\begin{figure}[h]
\centering
\parbox{10cm}{
\includegraphics[clip,scale=0.40]{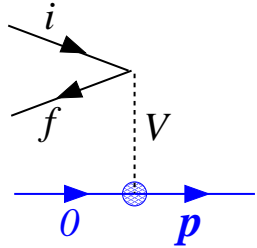}%{diagram3.eps}
}
\caption{
Decay of the atomic vacancy $i\to f$ via the fullerene ionization 
$\textcolor{blue}{0\to \bfp}$.
$V$ stands for the Coulomb atom-fullerene interaction, the filled 
circle denotes 
the effective vertex of the particle-hole excitation in the fullerene.
}
\label{diagram3.fig}
\end{figure}

In addition to the modification of the atomic Auger decay via the
change of the electric field at the atom, the presence of the fullerene
shell opens another channel of the Auger decay, which is absent in the
isolated atom 
\cite{CederbaumEtAl1997,AverbukhCederbaum2006,AmusiaBaltenkov2006}.
Diagrammatical representation of this process is given 
in figure \ref{diagram3.fig}.
The energy $\om_{if}$, released in the atomic transition $i\to f$, 
is transferred, by means of the Coulomb interaction, 
to the fullerene electron which becomes ionized.  
In \cite{AverbukhCederbaum2006} the 
process was called interatomic Coulomb decay, thus pointing out its 
similarity with the interatomic decay in smaller systems 
(e.g., in neon dimer \cite{SantraEtAl2000}). 

%%%%%%%%%%%%%%%%%%%%%%%%%% Figure fig4.eps

It was pointed out in \cite{AverbukhCederbaum2006,AmusiaBaltenkov2006}  
that this decay channel in the endohedral atom becomes especially important
if the non-radiative transition $i\to f$ is energetically 
forbidden in the isolated atom.  
In this case one compares the additional Auger decay width 
$\Gamma_{\nu_f\nu_i}^{\rm A@C_N}$ 
and the width of the radiative decay, $\Gamma_{fi}^{\gamma,\, \rm A}$.
The estimates, carried out in  
\cite{AverbukhCederbaum2006,AmusiaBaltenkov2006} for the atoms
placed at-the-center of the C$_{60}$ cage, show the dramatic increase
of the decay rate (by the factor 
$\Gamma_{\nu_f\nu_i}^{\rm A@C_N}/\Gamma_{fi}^{\gamma,\,\rm A}\sim 10^5-10^6$)
for the transitions with $\om_{fi}$ lying in vicinity of the
giant plasmon resonance at $20$ eV.
   
In this section we demonstrate, that for off-the-center atoms 
the increase of the decay rate can become even more pronounced and in much
wider range of the transition energies.

To construct the amplitude of the process presented in
figure \ref{diagram3.fig}, 
we use the approximate formula (\ref{atom-cage.4}) for the atom--fullerene
Coulomb interaction.
The amplitude is used further in (\ref{width1.1}) to derive the width.
Restricting ourselves to the dipole transitions $\nu_i\to \nu_f$,
we obtain the following expression for the 
partial width of the transition as a function of the atom displacement: 
\begin{eqnarray}
\Gamma_{\nu_f\nu_i}^{\rm dip,\, A@C_N}(a)
=
\calF(\om_{if};a)
\,
\Gamma_{\nu_f\nu_i}^{\gamma, {\rm A}}.
\label{additional2_4.5}
\end{eqnarray}
The factor $\calF(\om_{if};a)$ stands for the ratio of the Auger-decay width
in the endohedral atom to the radiative width in the isolated atom.
It depends on the displacement and the transition energy, and can be 
written as follows:
\begin{eqnarray}
\calF(\om_{if}; a)
=
{c^3 \over 2\om_{if}^3}\,
\sum_{l=1}^{\infty}
l\,(2l+1)\,a^{2l-2} \,
\Im\, \walpha_{l}^{{\rm C}_N}(\om_{if})\,.
\label{additional2_4.5a}
\end{eqnarray}
Similar to the series from (\ref{d_Auger_non_central5.13b}),
the expansion parameter is $(a/R)^2$.

Placing atom at the center and accounting for the 
approximate relation (\ref{Appendix_B.02}), one derives 
$\calF(\om_{if};0)=(3c^3/2\om_{if}^3)\,\Im\,\walpha_{1}^{{\rm C}_N}(\om_{if})
\approx (3c^3/2R^6\om_{if}^3)\,\Im\,\alpha_{1}^{{\rm C}_N}(\om_{if})$.
Using the latter expression in (\ref{additional2_4.5}), one arrives at
the formula derived in \cite{AmusiaBaltenkov2006}.

%%%%%%%%%%%%%%% figure fig5.eps
\begin{figure}[h]
\centering
\includegraphics[clip,scale=0.35]{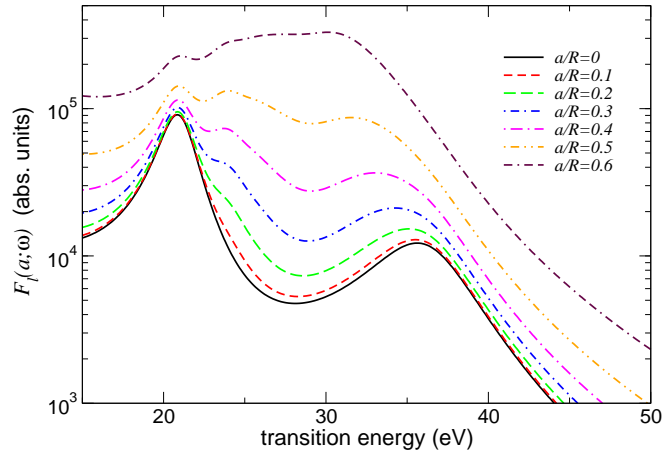}%{F_C80_g2_020.eps}
\caption{
Dependence of $\calF(a;\om)$ from (\ref{additional2_4.5a})
on the transition energy $\om$ calculated for several 
$a/R$ values as indicated.
The data refer to the fullerene C$_{80}^{6-}$
(see \ref{C80}) and explanations in the text.
}
\label{F_C80_beyond.fig}
\end{figure}
%%%%%%%%%%%%%%%%%%%%%%%%%% 
 
Dependence of the ratio $\calF(\om;a)$ on the transition energy, 
calculated for several $a/R$ values, is presented in figure 
\ref{F_C80_beyond.fig}.
The data refer to C$_{80}^{6-}$.
The parameters of the fullerene cage and of the dynamic polarizabilities 
$\walpha_{l}^{{\rm C}_N}(\om)$ are given in \ref{C80}. 

It is worth noting change in the $\calF(\om;a)$ profile with increase 
of the displacement $a$. 
For the at-the-center atom (or slightly off-the-center) the 
behaviour of $\calF(\om;a)$ is determined by the imaginary parts of 
$\walpha_{l}^{{\rm C}_N}(\om)$ with $l\approx 1$
 (see lower left graph in figure \ref{A_L_C80.fig}).
As a result, there is a pronounced maximum at $\approx 20$ eV, 
-- the energy of the first surface plasmon excitation.
In the vicinity of the second surface plasmon peak ($\approx 36$ eV)
the value of $\calF(\om;a)$ is less by an order of magnitude.
As the displacement increases, the higher-$l$ terms contribute to 
the right-hand side of (\ref{additional2_4.5a}). 
For $l\gg 1$ both plasmon modes merge, so that for each $l$ the 
dependence $\Im\,\alpha_{l}^{{\rm C}_N}(\om)$ has a single peak
at $\om \approx 30$ eV.
Hence, in this energy range the ratio $\calF(\om;a)$ acquires additional 
enhancement.
The curves with $a/R=0.4\dots 0.6$ show that large enhancement can be achieved 
in a wide interval of energies: $\om \approx 20\dots 40$ eV.

Therefore, we conclude that the decay channel, presented by figure
\ref{diagram3.fig}, can be additionally intensified (by more that the 
order of magnitude) if the endohedral atom is located sufficiently 
far away from the cage center.

%%%%%%%%%%%%%%%%%%%%%%%%%%%%%%%%%%%%%%%%%%%%%%%%
\subsection{The radiative decay in the endohedral atom
\label{Radiative}} 

\begin{figure}[h]
\centering
\includegraphics[clip,scale=0.40]{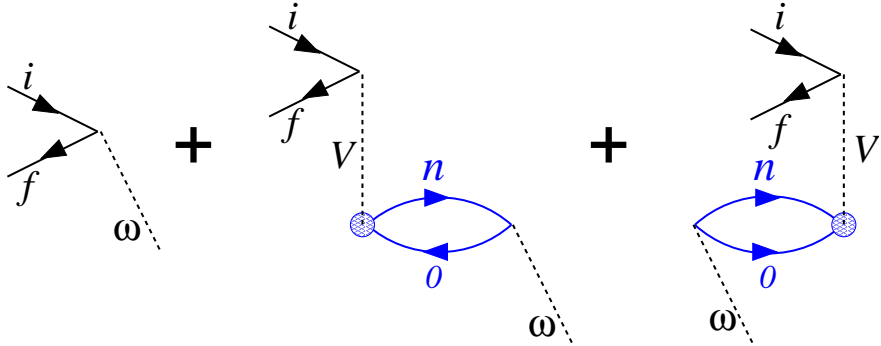}%{diagram4.eps}
\caption{
The radiative decay $i\to f$ in an isolated atom (the first diagram) and 
the correction due to the virtual excitation
$0\to n$ of the fullerene (two last diagrams).
The dashed line depicts the emitted photon, other notations as in 
figure \ref{diagram1.fig}.
The energy conservation law implies $\om = \om_{if}$.
}
\label{diagram4.fig}
\end{figure}

The radiative decay $i \to f + \gamma$ in an isolated atom
is represented by the first diagram in figure \ref{diagram4.fig}.
The correction due to the virtual polarization of fullerene under 
the joint action of the Coulomb interaction  and the photon field 
is illustrated by other two diagrams in  the figure.
For a centrally-positioned atom this process was considered in
 \cite{AmusiaBaltenkov2006}.

%%%%%%%%%%%%%%%%%%%%%%%%%% Figure fig6.eps

For a non-central position, applying the approximations described above, 
one derives the following formula which relates the partial widths of the
(dipole) radiative decay in the endohedral atom and in the isolated one:
\begin{eqnarray}
\Gamma_{\nu_f \nu_i}^{\gamma, {\rm A@C_{N}}}
=
%\Bigl|1-\beta^{\rm C_{N}}_1(\om_{if})\Bigr|^2
\calD(\om_{if})\,\Gamma_{\nu_f \nu_i}^{\gamma, {\rm A}}\,,
\label{Radiative4.1}
\end{eqnarray}
where
\begin{eqnarray}
\calD(\om_{if}) =
\Bigl|
1-\beta^{\rm C_{N}}_1(\om_{if})
\Bigr|^2\,.
\label{Radiative4.2}
\end{eqnarray}
Here $\beta^{\rm C_{N}}_1(\om_{if})$ stands for the dipole 
shielding factor of the fullerene.
This quantity determines the modification of the electric field 
{\it inside} the fullerene due to its polarization under the action
of the external uniform electric field 
 (see \ref{MultipolPolar} for more details).
Using the approximate relationship between $\beta^{\rm C_{N}}_1(\om_{if})$
and the dipole polarizability of the fullerene, 
$\beta^{\rm C_{N}}_1(\om_{if})\approx \alpha^{\rm C_{N}}_1(\om_{if})/R^3$
(see (\ref{Appendix_B.02})), one reduces the right-hand side of 
(\ref{Radiative4.1}) to the formula derived in 
 \cite{AmusiaBaltenkov2006} for the atom at-the-center.

The quantity $\calD(\om)$ is a so-called dynamical screening factor
\cite{SolovyovConnerade2005,LoKorolSolovyov2007,LoKorolSolovyov2009}.
It is equal to the ratio of the intensity of the total electric field 
$\bfE_{\rm tot}$ at the atom to the intensity of the external field $\bfE_0$ 
applied to the endohedral system $\rm A@C_N$. 
The total electric field $\bfE_{\rm tot}=\bfE_0+\Delta\bfE$ contains  
additional term $\Delta\bfE$ which is due to the shell polarization.
In the cited papers it was demonstrated that $\calD(\om)$ defines the 
increase of the photoabsorption rate of the atom due to the presence of the
fullerene shell.
Radiative decay is the inverse process for the photoabsorption. 
Therefore, the same dynamical screening factor   relates the
radiative width in the encaged and isolated atoms.

The factor $\calD(\om)$, as it is defined by (\ref{Radiative4.2}), does not 
depend on the distance $a$ from the cage center.
This is a consequence of the approximation made to derive equation 
(\ref{Radiative4.2}).
Namely, when calculating the additional electric field $\Delta\bfE$ we ignored
the reciprocal effect of the polarized atom on the shell.
Thus, it was assumed that the shell is polarized only under the action
of the {\it uniform} field $\bfE_0(\om)$ of the dipole photon. 
The field  $\bfE_0(\om)$ gives rise to a uniform polarization of the field
inside the cavity of the fullerene. 
As a result, the dynamical screening factor does is not sensitive to the 
position of the endohedral atom 
\cite{SolovyovConnerade2005,LoKorolSolovyov2007}.
Such an approximation must be modified if the atomic dynamic dipole 
polarizability $\alpha_1^{\rm A}(\om)$ is large enough to cause a 
noticeable additional polarization of the shell. 
In this case, as it was demonstrated in 
\cite{LoKorolSolovyov2009,Lo_Thesis}, the dynamical screening factor
acquires dependence on the displacement vector $\bfa$. 
In the present paper, for the sake of simplicity, we ignore this reciprocal
effect of the polarized atom on the shell.

%%%%%%%%%%%%%%% Figure fig7.eps
\begin{figure}[h]
\centering
\includegraphics[clip,scale=0.35]{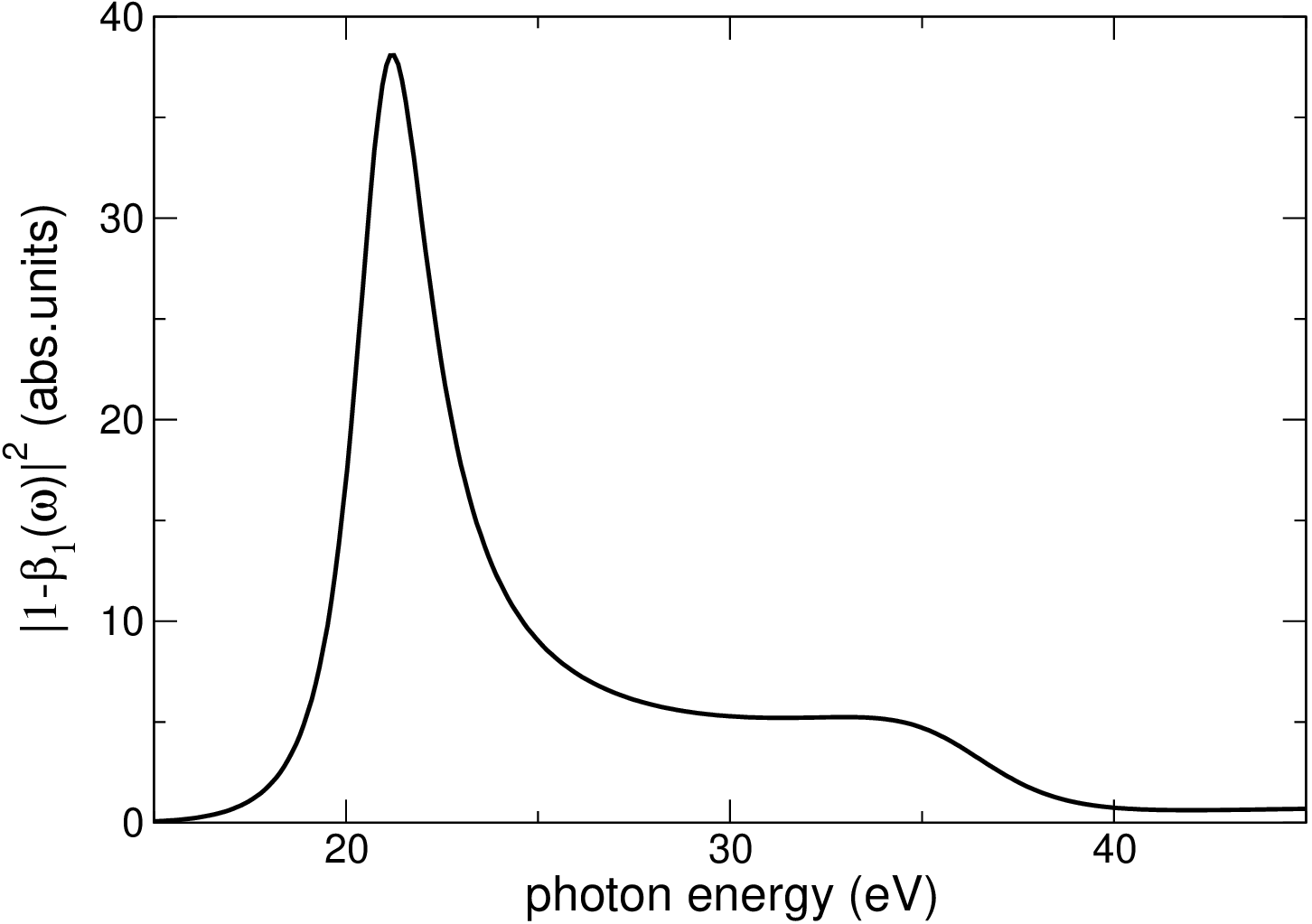}%{DynScren_C80_g2_020.eps}
\caption{
Dynamical screening factor 
$\calD(\om) =\Bigl|1-\beta^{\rm C_{N}}_1(\om)\Bigr|^2$
 for the fullerene C$_{80}^{6-}$.
}
\label{DynScren_C80.fig}
\end{figure}
%%%%%%%%%%%%%%%%%%%%%%%%%% 

The dependence of the dynamical screening factor on the photon
energy is presented in figure \ref{DynScren_C80.fig}.
The calculations, performed for C$_{80}^{6-}$, were carried out
within the plasmon resonance approximation, see \ref{PRA}.
The profile of this dependence, characterized by a powerful peak in
the vicinity of the symmetric surface plasmon energy and with the
extended right shoulder due to the activation of the antisymmetric
plasmon excitation, is similar to those reported earlier for other
spherically-symmetric fullerenes of finite thickness 
\cite{LoKorolSolovyov2007,Lo_Thesis}. 
Thus, one can expect the order-of-magnitude increase in the decay rate
magnitude for radiative transitions within $\approx 20\dots25$ energy range, 
and less than that for $\om \approx 25\dots 35$ eV. 

To conclude this section let us note that the applicability of the 
plasmon resonance approximation for the dynamical screening factor 
was tested in  \cite{LoKorolSolovyov2009,LoKorolSolovyov_ISACC2009,Lo_Thesis}
against the TDLDA calculations \cite{MadjetEtAl_2007,ChakrabortyEtAl_2008}.
A good quantitative agreement between these two approaches was reported.

%%%%%%%%%%%%%%%%%%%%%%%%%%%%%%%%%%%%%%%%%%%%%%%%
\section{Numerical results
\label{Numerics}} 

In this section we apply the approach described above to carry out 
model calculations of the photoionization cross section of 
the endohedral complex Sc$_3${\rm N@C}$_{80}^{+}$ in the photon energy
range $30\dots 45$ eV.
Within this range of photon energies the photoionization spectrum 
of an isolated ion Sc$^{2+}$ is dominated by autoionizing resonances 
(of the widths less than 1 eV) due to $3p$ excitations
\cite{SchippersEtAl_2003,AltunManson_1999,SossahZhouManson2008}.
However, recent experiments by  M\"{u}ller \etal \cite{MuellerEtAl2007}
demonstrated that a distinct resonance structure is absent in the
photoionization spectrum of the ion Sc$_3${\rm N@C}$_{80}^{+}$.

In what follows, we start with the quantitative analysis of the 
dependence of the autoionizing widths in the encaged Sc$^{2+}$
on the displacement of the ion from the cage center due to the 
mechanisms discussed in sections \ref{Problem} and \ref{additional}.
The obtained data are used further to calculate the photoionization
spectrum of the endohedral complex.

%%%%%%%%%%%%%%%%%%%%%%%%%%%%%%%%%%%%%%%%%%%%%%%%
\subsection{Auger widths in Sc$^{2+}$ and in ${\rm Sc}^{2+}{\rm @C}_{80}^{5-}$
\label{Numerics_Sc2+}}

Cross sections for the 
photoionization (PI) of Sc$^{2+}$ ions were measured by employing 
the merged ion-photon beams technique \cite{SchippersEtAl_2003}.
It was noted (see also \cite{SossahZhouManson2008}), that in addition 
to a direct photoionization pathway, 
the process can proceed via an intermediate resonance state 
according to the scheme 
\begin{eqnarray}
\hbar\om + \mbox{Sc}^{2+}\longrightarrow
\left(\mbox{Sc}^{2+}\right)^{*}
\longrightarrow
\mbox{Sc}^{3+} + \ee^{-}\,.
\label{Numerics_Sc2+.1}
\end{eqnarray}
The intermediate state decays via (super-)Coster-Kronig transition resulting in
a resonance line in the PI spectrum. 

The Sc$^{2+}$ ion beam, used in the experiment, contained  
ions in the ground state 
(the configuration $[{\rm Ne}]3s^23p^6 3d\, ^2D_{3/2}$) and in the 
first two excited, metastable states 
($[{\rm Ne}]3s^23p^6 3d\, ^2D_{5/2}$ and $[{\rm Ne}]3s^23p^6 4s\, ^2S_{1/2}$)
as well. 
This lead to a variety of the intermediate states 
$\left(\mbox{Sc}^{2+}\right)^{*}$ and, as a result, to a number of 
measured peaks in the spectrum within the photon energy range 
$30\dots 45$ eV.
The parameters of experimentally measured peaks is contained
in tables and figures in \cite{SchippersEtAl_2003}. 
Theoretical data on the resonances together with the 
theory-versus-experiment discussion 
 one finds in \cite{SossahZhouManson2008}
(see also \cite{Kjeldsen2006}).

For our case study we restrict ourselves to the transitions from 
the ground state only. 
In table \ref{Sc2+_other1.table1} the data are presented on most 
pronounced PI resonances within the interval $\om=31\dots 42$ eV.
For each excited state $\left(\mbox{Sc}^{2+}\right)^{*}$
the following information is included:
resonance energy $\om_{21}$
(subscripts "1" and "2" mark the ground and the excited states, 
respectively), 
peak value $\sigma$ of the PI cross section, 
Auger width $\Gamma^{({\rm Sc}^{2+})^*}$, 
peak area $\calS$, 
radiative width $\Gamma_{2\to 1}^{\gamma}$ 
and the parameter 
$\kappa=
\Bigl(\Gamma^{({\rm Sc}^{2+})^*}\sigma/
\Gamma_{2\to 1}^{\gamma}\Bigr)^{1/2}$ 
calculated in accordance with  (\ref{d_Auger_non_central5.13a}).

The values of $\om_{21}$, $\sigma$, $\Gamma^{({\rm Sc}^{2+})^*}$
and $\calS$ were deduced from the data presented in
Table II and Figs. 4,5,8-10 in \cite{SchippersEtAl_2003}.
When indicated, the Auger width (or the cross section $\sigma$) 
were calculated assuming the  Lorentz profile of the peak.
This was done using the relation 
$\sigma\,\Gamma^{({\rm Sc}^{2+})^*} = 2\calS/\pi$. 
Thus, in our calculations we ignored the asymmetry seen in 
the experimentally measured peaks.

%%%%%%%%%%%%%%%%%%%%%%%%%%%%%
\begin{table}[h]
\centering
\caption{
Transition energies $\om_{21}$ from the excited state 
$\left(\mbox{Sc}^{2+}\right)^{*}$ to the ground state
$\mbox{Sc}^{2+}(D_{3/2})$, 
peak widths $\Gamma^{({\rm Sc}^{2+})^*}$,
peak values of the PI cross section $\sigma$,
peak areas $\calS$,
radiative widths $\Gamma_{2\to 1}^{\gamma}$
and parameters $\kappa$ (see equation (\ref{d_Auger_non_central5.13a})).
\\}
\begin{tabular}{|r|c|c|c|c|c|c|}
\hline
     &      &       &          &      &      &       
\\
  $\left(\mbox{Sc}^{2+}\right)^{*\,{\rm a}}$ 
               & $\om_{21}$  
               & $\Gamma^{({\rm Sc}^{2+})^*}$ 
               & $\calS$  
               & $\sigma$ 
               & $\Gamma_{2\to 1}^{\gamma}$  
               & $\kappa$  
\\
     & (eV) & (meV) & (eV\,Mb) & (Mb) & ($10^{-6}$ eV) & 
\\
\hline
$3d^2\,^2\rmF_{5/2}$& 31.66& 116 &  0.87   &  4.8$^{\rm b}$     
 & 0.15 & $4.71\times 10^{-4}$\\
$3d4s\,^2\rmP_{1/2}$& 33.22&  45 &  1.22   & 17.3$^{\rm b}$     
 & 0.70 & $3.10\times 10^{-3}$\\
$3d4s\,^2\rmF_{5/2}$& 34.73&  53 &  5.40   &  64.9$^{\rm b}$    
 & 1.13 & $7.02\times 10^{-3}$\\
$3d^2\,^2\rmF_{5/2}$& 37.14& 847 &  47.9   &  36.0              
 & 11.4 & $4.17\times 10^{-3}$\\
$3d^2\,^2\rmP_{1/2}$& 39.63& 11.5$^{\rm c}$& 10.1& 560$^{\rm c}$
 & 8.24 & 0.120 \\
$3d^2\,^2\rmP_{3/2}$& 39.72& 11.3$^{\rm d}$& 1.60&  90$^{\rm c}$
 & 0.65 & $1.36\times 10^{-2}$ \\
$3d^2\,^2\rmD_{5/2}$& 40.22&  2.3$^{\rm d}$& 1.60& 450$^{\rm c}$
 & 0.45 & $5.60\times 10^{-2}$ \\
\hline
\hline
\end{tabular}

\vspace*{0.2cm}
\parbox{13cm}{
 $^{\rm a}$ {\footnotesize $[{\rm Ne}]3s^23p^5$ is omitted in the 
excited-state designation.}
\\
$^{\rm b}${\footnotesize Calculated from the Lorenz profile
(see explanation in the text)}.
\\
$^{\rm c}${\footnotesize Deduced  by digitalizing Figs. 9 and 10 in
\cite{SchippersEtAl_2003}.}
}
\label{Sc2+_other1.table1}
\end{table}
%%%%%%%%

The radiative width of the transition $2\to 1$ was calculated as 
follows
\begin{eqnarray}
\Gamma_{2\to 1}^{\gamma}
=
{2\om_{21}^2 \over c^3}\,
{g_1 \over g_2}\, f_{1\to 2}
\label{RadWidth2.2}
\end{eqnarray}
where $g_1=4$ and  $g_2$ are statistical weights of the ground and the
excited states, and $f_{1\to 2}$ is the oscillator strength of the
transition $1\to 2$.
For a given transition $f_{1\to 2}$ is related to the 
PI  peak area $\calS$ through
$f_{1\to 2} = (c/ 2\pi^2) \, \calS$.

Once the values of $\Gamma_{2\to 1}^{\gamma}$  and  $\kappa$ are established
one can use them in 
 (\ref{d_Auger_non_central5.11}), (\ref{d_Auger_non_central5.12}), 
(\ref{additional2_4.5}) and (\ref{Radiative4.1}) 
to calculate the total widths of the excited states
for the ion  ${\rm Sc}^{2+}$  encaged in a fullerene.
The results of such calculations, 
carried out for the system  ${\rm Sc}^{2+}@{\rm C}_{80}^{5-}$,
are presented in table \ref{table.total_width_1}.

%%%%%%%%%%%%%%%%%%%%%%%%%%%%%
\begin{table}[h]
\centering
\caption{
Total widths $\Gamma_{\rm tot}^{({\rm Sc}^{2+})^{*}@{\rm C}_{80}^{5-}}(a)$
 for several $a/R$ ratios.
For each transition the upper line is due to the 
$\Gamma_{2\to 1}^{\gamma}$ and $\kappa$
data from table \ref{Sc2+_other1.table1},
the lower line was obtained accounting for the correction
due to Sossah \etal \cite{SossahZhouManson2008} (see explanation in the text).
\\}
\begin{tabular}{|l|c|c|ccccc|}
\hline
    $\left(\mbox{Sc}^{2+}\right)^{*}$ 
    & $\om_{21}$
    & $\Gamma^{({\rm Sc}^{2+})^{*}}$ 
    &\multicolumn{5}{|c|}{
$\Gamma_{\rm tot}^{({\rm Sc}^{2+})^{*}@{\rm C}_{80}^{5-}}(a)$ in eV}
\\
             &    (eV)   &      (meV) 
             &{\small$a/R=0$} 
             &{\small $a/R=0.4$}
             &{\small $a/R=0.5$}
             &{\small $a/R=0.55$}
             & {\small$a/R=0.6$} 
\\
\hline
$3d^2\,^2\rmF_{5/2}$ & 31.66 & 116 & 0.116 & 0.118 & 0.127 & 0.141 & 0.177\\
                     &       &     & 0.115 & 0.120 & 0.135 & 0.158 & 0.223\\
\hline
$3d4s\,^2\rmP_{1/2}$ & 33.22 & 45  & 0.048 & 0.071 & 0.117 & 0.183 & 0.351 \\ 
                     &       &     & 0.049 & 0.090 & 0.175 & 0.302 & 0.650 \\
\hline
$3d4s\,^2\rmF_{5/2}$ & 34.73 & 53  & 0.062 & 0.114 & 0.215 & 0.361 & 0.761\\ 
                     &       &     & 0.069 & 0.165 & 0.370 & 0.685 & 1.595\\
\hline
$3d^2\,^2\rmF_{5/2}$ & 37.14 & 847 & 1.027 & 1.343 & 1.829 & 2.432 & 3.849\\
                     &       &     & 1.150 & 1.702 & 2.599 & 3.768 & 6.678\\
\hline
$3d^2\,^2\rmP_{1/2}$ & 39.63 & 11.5& 0.118 & 0.363 & 0.931 & 1.875 & 4.769\\ 
                     &       &     & 0.246 & 0.837 & 2.267 & 4.685 & 12.19\\
\hline
$3d^2\,^2\rmP_{3/2}$ & 39.72 & 11.3& 0.018 & 0.028 & 0.044 & 0.066 & 0.122\\ 
                     &       &     & 0.022 & 0.041 & 0.073 & 0.119 & 0.243\\
\hline
$3d^2\,^2\rmD_{5/2}$ & 40.22 & 2.3 & 0.008 & 0.020 & 0.044 & 0.082 & 0.195\\ 
                     &       &     & 0.013 & 0.038 & 0.094 & 0.186 & 0.468\\
\hline
\end{tabular}
\label{table.total_width_1}
\end{table}
%%%%%%%%%%%%%%%%%%%%%%%%%%%%%

The first three columns in table \ref{table.total_width_1} are identical to 
those from table \ref{Sc2+_other1.table1} and are reproduced for the sake 
of convenience only.
The presented values of the total width 
$\Gamma_{\rm tot}^{({\rm Sc}^{2+})^*@{\rm C}_{80}^{5-}}(a)$ were obtained 
by summing three terms, calculated from (\ref{d_Auger_non_central5.11}), 
 (\ref{additional2_4.5}) and (\ref{Radiative4.1}) for several values of the 
$a/R$ ratio  as indicated. \footnote{The contribution of the modified 
radiative width, equation ((\ref{Radiative4.1}), to  
$\Gamma_{\rm tot}^{({\rm Sc}^{2+})^*@{\rm C}_{80}^{5-}}(a)$
is negligent compared with the other two terms. It is mentioned for the 
sake of  consistency only.}

For each excited state  $\left(\mbox{Sc}^{2+}\right)^{*}$ the upper line 
presents the values of 
$\Gamma_{\rm tot}^{({\rm Sc}^{2+})^*@{\rm C}_{80}^{5-}}(a)$
obtained using the data on $\Gamma_{2\to 1}^{\gamma}$ and $\kappa$ 
from table \ref{Sc2+_other1.table1}.
The lower line was obtained using the $\Gamma_{2\to 1}^{\gamma}$ and 
$\kappa$ values corrected due to the recommendation made by 
Sossah \etal \cite{SossahZhouManson2008}.

In the cited paper theoretical calculations of the 
PI cross section of Sc$^{2+}$ were performed for photon energies from 
threshold to 68.0 eV.
The result of calculations were compared to the experimental data from 
\cite{SchippersEtAl_2003}.
One of the results  of this comparison concerns the discrepancy in the
calculated and measured values of the total oscillator strength of the 
3$p^6$ shell, which must be equal to 6 in accordance with the 
Thomas-Reiche-Kuhn sum rule.
On the basis of physical arguments, presented in pp. 7 and 8 of 
\cite{SossahZhouManson2008}, the authors indicate that
"\dots the total oscillator strength in the 
photoionization cross section of Sc$^{2+}$ from threshold to 68 eV
should be a bit under 6\dots".
They mentioned further, that the calculated total oscillator strength 
is 5.29, whereas the one which follows from the experimental data equals 
to 3.24, which is too low.
The conclusion, which is drawn by the authors, is as follows: 
"Thus we believe that the overall magnitude of the measured cross section
is too small and should be multiplied by a factor of 5.29/3.24=1.63
to bring the oscillator strength to a reasonable value." 

One easily verifies that such multiplication increases
$\Gamma_{2\to 1}^{\gamma}$ and $\kappa$ by the 
same factor of 1.63. 
The lower lines in table \ref{table.total_width_1} were obtained using these
enhanced values.

The data in table \ref{table.total_width_1} illustrate the sensitivity of 
the width $\Gamma_{\rm tot}^{({\rm Sc}^{2+})^{*}@{\rm C}_{80}^{5-}}(a)$ both
on the atom displacement and on the excited state.
While for all excited states $\left(\mbox{Sc}^{2+}\right)^{*}$ the widths
increase with $a$, the increase rate is quite different for different states.
It can be quite moderate as $a/R$ varies from 0 to 0.6
(e.g., the increase by approximately 50 \%  in the case of 
$3d^2\,^2\rmF_{5/2}$  at $\om_{21}=31.66$ eV) or very sharp
(up to $10^3$ increase for the $3d^2\,^2\rmP_{1/2}$ state). 

Let us briefly describe the relative contribution 
of the  two mechanisms, presented by figures \ref{diagram1.fig}(b) and 
\ref{diagram3.fig}, to the total width 
$\Gamma_{\rm tot}^{ ({\rm Sc}^{2+})^{*}@{\rm C}_{80}^{5-}}(a)$.
With the exception for the  $3d4s\,^2\rmP_{1/2}$ excited state, 
the contribution due to the diagram \ref{diagram1.fig}(b) dominates by the 
factor from 2 to 20 depending on the state and on the $a/R$ value.
In the case of $3d4s\,^2\rmP_{1/2}$ both channels contribute to the total 
width almost equally.

%%%%%%%%%%%%%%%  Figure fig8.eps
\begin{figure}[h]
\centering
\includegraphics[clip,scale=0.35]{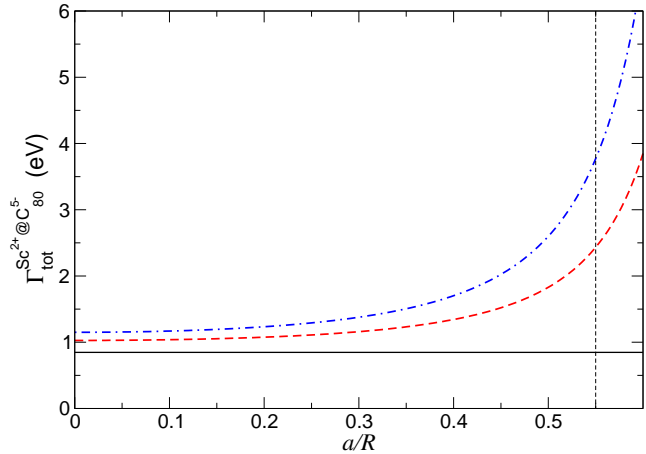}%{width_3d2_om37_g2_020.eps}
\caption{
Total width $\Gamma_{\rm tot}^{ ({\rm Sc}^{2+})^{*}@{\rm C}_{80}^{5-}}(a)$
of the excited $3d^2\, ^2{\rm F}$ state (excitation energy 
$\om_{21}=37.14$ eV) versus $a/R$.
The dashed curve was calculated using the $\Gamma_{2\to 1}^{\gamma}$ and 
$\kappa$ data from table \ref{Sc2+_other1.table1}.
The chained curve accounts for the 1.63-correction due to 
Sossah \etal \cite{SossahZhouManson2008} (see explanation in the text).
Horizontal solid line marks the width in the isolated ion Sc$^{2+}$.
The vertical line marks the ratio $a/R=0.55$ which can be estimated as
the largest available for Sc$^{2+}$ in Sc$_3$N@C$_{80}$
(see section \ref{Numerics_Sc3N@C80}).
}
\label{Width_3d2_F.fig}
\end{figure}
%%%%%%%%%%%%%%%%%%%%%%%%%% 

In figure \ref{Width_3d2_F.fig} the dependence 
$\Gamma_{\rm tot}^{({\rm Sc}^{2+})^{*}@{\rm C}_{80}^{5-}}(a)$ 
is plotted for the excited $3d^2\, ^2{\rm F}$ state (excitation energy 
$\om_{21}=37.14$ eV) which is responsible for the widest peak in the PI 
cross section from the ground state of Sc$^{2+}$ \cite{SchippersEtAl_2003}. 
In the encaged ion the width can attain noticeably larger values than
in the isolated ion (the horizontal line in the graph), and 
this influences the profile of the PI spectrum.

To analyze the influence of the fullerene cage on the PI spectrum 
of $\rmSc^{2+}$ one can use the following simple model.
The cross section $\sigma_{{\rmSc}^{2+}}(\om;a)$ 
is calculated as the sum of resonance terms corresponding to the
intermediate excited states  $\left(\mbox{Sc}^{2+}\right)^{*}$ 
listed in tables  \ref{Sc2+_other1.table1} and \ref{table.total_width_1}.
Each term is approximated by a symmetric Lorentzian line profile.
Thus, the model does not account for the asymmetry of the 
peaks (see the discussion in \cite{SchippersEtAl_2003}).
Hence, $\sigma_{{\rmSc}^{2+}}(\om;a)$ was written as follows:
\begin{eqnarray}
\sigma_{{\rmSc}^{2+}}(\om;a)
=
{1\over 2\pi}\sum_{j=1}^7
{ \calS_j \Gamma_j \over (\om -\om_j)^2 + \Gamma_j^2/4}\,,
\label{Numerics_Sc3N@C80.2}
\end{eqnarray}
where $j$ enumerates the excited states.
For each $\left(\mbox{Sc}^{2+}\right)^{*}$ the values of resonance frequency 
$\om_j$, peak area $\calS_j$ and peak width
(for the isolated ion as well as for the encaged one) 
one finds in the tables.

%%%%%%%%%%%%%%%%% Figure fig9.eps
\begin{figure}[h]
\centering
\includegraphics[clip,scale=0.4]{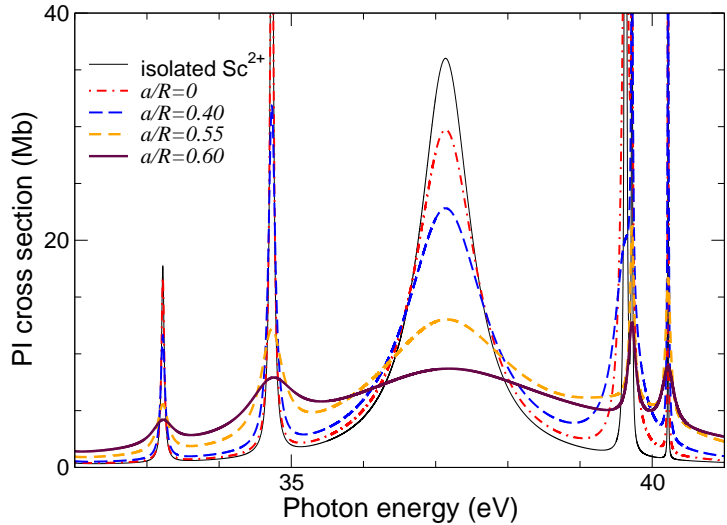}%{Shippers_profiles_020.eps}
\caption{
The PI cross section of the isolated $\rmSc^{2+}$ ion (thin solid line)
and of the $\rmSc^{2+}$ ione encaged in $\rmC_{80}^{5-}$ 
at different distances from the center (as indicated). 
}
\label{Shippers_profiles_020.eps}
\end{figure}
%%%%%%%%%%%%%%%%%%%%%%%%%% 

Figure \ref{Shippers_profiles_020.eps} presents the dependence of  
$\sigma_{{\rmSc}^{2+}}(\om;a)$ on $\om$ calculated 
for the isolated scandium ion (thin solid curve) and for $\rmSc^{2+}$ 
encaged in  $\rmC_{80}^{5-}$ at different distances from the center.
To calculate the resonance terms in (\ref{Numerics_Sc3N@C80.2}) 
we used,  for each indicated value of the ratio $a/R$, 
$\Gamma_{\rm tot}^{({\rm Sc}^{2+})^{*}@{\rm C}_{80}^{5-}}(a)$ listed 
in the upper lines in table \ref{table.total_width_1}.
Within the model framework we assumed the peak areas $\calS$ 
(see the third  column in table \ref{Sc2+_other1.table1})
to be independent on $a$.
Also, we disregarded possible redistribution of the ionic oscillator 
strength due to the presence of the fullerene shell \cite{MuellerEtAl2008}.

The focus of our study is on the modification of the PI spectrum profile
due to the Auger widths broadening.
From this viewpoint the comparison of different curves in figure
\ref{Shippers_profiles_020.eps} is quite illustrative.
For moderate values of $a/R$, similar to the case of the isolated ion, the
spectrum is dominated by well-separated lines.
As $a$ increases, the lines, loosing in the peak values and gaining in
the widths,  become less accented. 
For $a/R = 0.5\dots 0.6$ (this range of the $a/R$ ratio  is quite realistic, 
see section \ref{Numerics_Sc3N@C80}) the spectrum  noticeably flattens. 
This effect is most pronounced for $\om \lesssim 35\dots 37$ eV,
where the effect of the fullerene shell polarization on the ionic 
Auger decay is very strong 
(see figures \ref{Sigma_C80.fig} and \ref{F_C80_beyond.fig}).

%%%%%%%%%%%%%%%%%%%%%%%%%%%%%%%%%%%%%%%%%%%%%%%%
\subsection{Photoionization spectrum of $\rmSc_3\rmN @\rmC_{80}^{+}$
\label{Numerics_Sc3N@C80}}

In \cite{MuellerEtAl2007} the experimental results for the PI 
cross section were reported for $\rmSc_3\rmN @\rmC_{80}^{+}$.
To carry out theoretical investigation of the PI process of this 
endohedral system  we assume that both its electronic structure and isomeric
form are close to those of the neutral compound $\rmSc_3\rmN @\rmC_{80}$.
which is a member of a family of the tri-metallic-nitride endohedral 
system \cite{DunschYang2007}.
These objects are rather curious in that often both of the two subsystems 
are unstable in isolation. 
When these two subsystems are combined into an endohedral fullerene, 
there is a series of electron transfers between various components of 
the subsystem that results in a mutual stabilizing effect 
\cite{DunschEtAl2004}.

Electronic structure of $\rmSc_3\rmN @\rmC_{80}$ metallofullerene 
was discussed in \cite{AlvarezEtAl_2002}. 
Each scandium atom donates 2 electrons to the fullerene cage. 
Additionally, a partial charge of 0.4 is donated to the nitrogen atom.	
The endohedral complex is therefore 
$(\rmSc^{2.4+})_3\rmN^{1.2+}@\rmC_{80}^{6-}$.
For the purposes of the present paper we ignore the partial charges
in all intermediate stages of the calculations. 
Hence,  each scandium ion is treated as doubly ionized, $\rmSc^{2+}$,
and the nitrogen is considered as a neutral atom.
At the final stage, when constructing the PI cross section of the
compound $\rmSc_3\rmN @\rmC_{80}^{+}$ the partial charges will be 
accounted for following the phenomenological arguments presented in 
\cite{MuellerEtAl2007}.

The isomeric form of the $\rmC_{80}^{6-}$ cage becomes of the $I_h$ type, i.e. 
'nearly spherical'.
Basing on the study of Nakao \etal \cite{NakaoKutitaFujita_1994} (see also
\cite{BaowanThamwattanaHill_2007})  one can use the value  $R=4.15$ \AA\, for
the average radius of the isomer.

All four atoms of the $\rmSc_3\rmN$ complex lay in a single plane 
\cite{StevensonEtAl1999,SlaninaNagase2005}.
The nitrogen atom forms the central part of the structure, and
the three scandium atoms are positioned at an average separation 
of 2.0 \AA.
Therefore, had the nitrogen atom resided at the cage center the
(average) displacement of each $\rmSc^{2+}$ would be $a=2$ \AA,
i.e., $a/R\approx 0.5$.
However, analyzing the NMR spectra of $\rmSc_3\rmN @\rmC_{80}$, 
Stevenson \etal \cite{StevensonEtAl1999} indicated that the complex is 
not localized in any particular site in the fullerene.
Therefore, it is meaningful to estimate the largest displacement 
$a_{\max}$ which can be experienced by $\rmSc^{2+}$ in the $\rmC_{80}^{6-}$ 
cage.
Calculating the density of electron cloud in the ground state of 
$\rmSc^{2+}$ (for example, by means of the Hartree-Fock code
\cite{ChernyshevaYakhontov_CPC}), one finds that
$R_{\rmSc^{2+}} \approx 1.1$ \AA\, is a good estimate for the radius of 
the ion.
Taking into account that the thickness of the fullerene electron cloud 
is  1.5 \AA\, \cite{RuedelEtAl2002} for C$_{60}$, 
we find the shortest distance between $\rmSc^{2+}$  and the cage
as $1.1+0.75=1.85$ \AA.
This value suggests that the largest displacement of Sc$^{2+}$ from the 
center is $a_{\max} \approx R - 1.85\, \mbox{\AA} = 2.3 \, \mbox{\AA}$,
yielding $a_{\max}/R \approx 0.55$.

Hence, to estimate the increase in the Auger widths in Sc$^{2+}$ due 
to the excitation of the plasmon oscillations in the fullerene shell 
one can use the $a/R$ ration from the range $0.5\dots 0.55$
(see table \ref{table.total_width_1} and figures 
\ref{Width_3d2_F.fig} and \ref{Shippers_profiles_020.eps}).

In the photon  energy range $30\dots 45$ eV the PI cross sections of the 
fullerene cage \cite{MuellerEtAl2007} and of the nitrogen atom 
\cite{SamsonAndel1990} are smooth functions of $\om$.
Therefore, one would expect that any peculiarity in the cross section 
$\sigma_{{\rm Sc}_3{\rm N@C}_{80}^{+}}$ of the endohedral complex is due 
to the scandium ions.
Ignoring the difference in the PI cross section between the $\rmC_{80}^{+}$ 
and $\rmC_{80}^{5-}$ 
cages\footnote{Such an approximation seems reasonable, since 6 extra valence 
electrons do not affect noticeably the dynamics of the surface plasmons which
define the magnitude and behaviour of the cross section in the indicated
$\om$-interval.}
one can write the following approximate relation:
\begin{eqnarray}
\Delta \sigma(\om;a)
\equiv
\sigma_{{\rm Sc}_3{\rm N@C}_{80}^{+}}
-
\sigma_{\rmC_{80}^{+}}(\om)
\approx
\sigma_{\rmN}(\om)
+ 
3
\sigma_{{\rmSc}^{2+}}(\om;a)\,.
\label{Numerics_Sc3N@C80.1}
\end{eqnarray}
The experimental data on $\sigma_{{\rm Sc}_3{\rm N@C}_{80}^{+}}$ and
$\sigma_{\rmC_{80}^{+}}(\om)$ one finds in figure 3 in 
\cite{MuellerEtAl2007}.
Therefore, to carry out theory-versus-experiment comparison it is
necessary to evaluate the right-hand side of (\ref{Numerics_Sc3N@C80.1}).
To calculate the term $\sigma_{\rmN}(\om)$ we used the tabulated data
\cite{HenkeData}.
The calculation of $\sigma_{{\rmSc}^{2+}}(\om;a)$ 
was carried out in accordance with (\ref{Numerics_Sc3N@C80.2}).
Additionally, the obtained $\sigma_{{\rmSc}^{2+}}(\om;a)$ curves were 
shifted on the photon energy axis by $+1.5$ eV.
This shift accounts (approximately) for the increase in $3p$ subshell 
binding energy  of the  Sc  charge state  $2.4+$ instead of $2+$ 
\cite{MuellerEtAl2007}.

%%%%%%%%%%%%%%%%%%%%%%%%%%%%  Figure with Cross Sections
%%%%%%%%%%%%%%%%%%%% Figure fig10.eps
\begin{figure}[h]
\centering
\includegraphics[clip,scale=0.45]{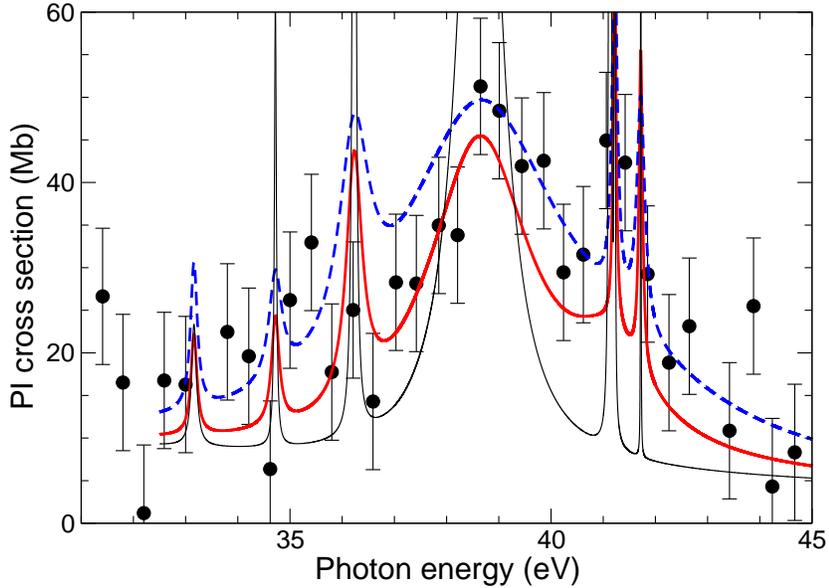}%{055_Manson_Shipp_Exp_020.eps}
\caption{
Theoretical versus experimental data on 
$\Delta \sigma(\om;a)$ (see equation (\ref{Numerics_Sc3N@C80.1})).
{\it Closed circles with error bars}: digitalized data from
figure 3 in \cite{MuellerEtAl2007}.
{\it Thin solid curve:} cross section $\Delta \sigma(\om;a)$
with the term $3\sigma_{{\rmSc}^{2+}}(\om;a)$ 
calculated for the isolated Sc$^{2+}$ ion.
{\it Thick curves:} calculations carried out for Sc$^{2+}$ ions displaced 
by $a=0.55 R= 2.3$ \AA\, from the cage center;
the {\it solid} curve  
obtained without the 1.63-correction due to \cite{SossahZhouManson2008},
the {\it dashed} -- the correction was accounted for.
See explanation in the text.
}
\label{Sc3N@C80.fig}
\end{figure}
%%%%%%%%%%%%%%%%%%%%%%%%%% 
%%%%%%%%%%%%%%%%%%%%%%%%%%%%%%%%%%%%%%%%%%%%%%%%%%%%%%%%
In figure \ref{Sc3N@C80.fig} we present the calculated dependences
$\Delta \sigma(\om;a)$  together with the experimental data
which were obtained by digitalizing the upper graph from
figure 3 in \cite{MuellerEtAl2007}.
The fitting, carried out in the cited paper, suggests that 
the experimental data can be described by a single broad peak of 6 eV
full width at half maximum (this profile is not drawn in figure
\ref{Sc3N@C80.fig}).
This behaviour clearly differs from that of the thin solid curve, which
was obtained by using the PI cross sections for the isolated 
$\rmSc^{2+}$ ion in  (\ref{Numerics_Sc3N@C80.1}).

The correspondence between theory and experiment can be considerably 
improved by accounting for the shell polarization and placing the 
scandium ions away from the cage center.
Two thick curves in figure \ref{Sc3N@C80.fig}
represent $\Delta \sigma(\om;a)$ calculated for $a =0.55 R= 2.3$ \AA.
The solid curve was obtained by using the $\calS$ values from 
table \ref{Sc2+_other1.table1} and the upper values of
$\Gamma_{\rm tot}^{({\rm Sc}^{2+})^{*}@{\rm C}_{80}^{5-}}(a)$ from
table \ref{table.total_width_1}.
The dashed curve was obtained by accounting for the correction
due to Sossah \etal \cite{SossahZhouManson2008}.
Within this scheme the peak areas $\calS$ were multiplied by 1.63 
and the lower values from the  $a/R=0.55$ column
in table \ref{table.total_width_1} were used.

Taking into account the approximate framework used to describe
the dynamic polarizational effects in the fullerene shell (the resonance
plasmon approximation accompanied by several parameters introduced 
"by hand") as well as several simplifying and phenomenological 
assumptions made to calculate the right-hand side of  
(\ref{Numerics_Sc3N@C80.1}), 
we would not claim that these curves reproduce in full 
the experimental data.
The main drawback of the calculated profiles is that they still contain
several peaks associated with the ionic transitions indicated in table
 \ref{Sc2+_other1.table1}. 
These features has not been seen in the experiment 
\cite{MuellerSchippers_2010}. 
Nevertheless, the general trend, which was intuitively formulated in 
\cite{MuellerEtAl2007}, is quite clear: 
the presence of the fullerene cage could result in significant broadening
of atomic/ionic resonances in photoionization.
The case study, which has been carried in this paper and which is illustrated
by  figure  \ref{Sc3N@C80.fig}, indicates that the impact of the cage 
polarization on the PI peaks strongly depends on (a) the location of the 
target atom inside the cage, (b) the energy of the atomic transition,
(c) the magnitudes of the radiative and non-radiative width of the transition.

%Two other (thick) curves in the figure present  $\Delta \sigma(\om)$
%calculated with the broadened peaks in the cross section
%$\sigma_{{\rmSc}^{2+}}(\om;a)$.
%Both curves were calculated for the case when all three Sc$^{2+}$ ions are 
%displaced by $a=0.55 R= 2.3$ \AA\ from the cage center. 
%The solid curve was obtained by using the peak vaules $$
% (\ref{Numerics_Sc3N@C80.2}).

%%%%%%%%%%%%%%%%%%%%%%%%%%%%%%%%%%%%%%%%%%%%%%%%%%%%%%%%%
\section{Summary
\label{Summary}} 

In summary, we have demonstrated that the Auger decay rate in an 
endohedral atom is very sensitive 
to the atom's location in the fullerene cage
as well as 
to the energy $\om$ released in the decay process.

Two additional decay channels have been considered, 
which appear in an endohedral system and
lead to the modification of the rate.
Firstly, there is a correction to the direct Auger-decay channel due to the
change in the electric field at the atom caused by dynamic polarization 
of the fullerene electron shell by the Coulomb field of the decaying vacancy.
In principle, this correction can be of either sign (depending on $\om$),
and, thus, can either increase or decrease the decay rate.
Within the second channel the released energy is transferred to the 
fullerene electron via the Coulomb interaction.

The contributions of both additional mechanisms are expressed in terms of 
the fullerene dynamic multipole polarizabilities of the second kind.
This type of polarizabilities appears when one is interested in the 
modification of the electric field in the interior of any hollow object (e.g., a fullerene)
due to its polarization under the action of the external field whose
source is also located inside  the object.
This is in contrast to the "conventional" polarizability
which is responsible for the same effect but in the case, when both 
the source and the observation point are located in the exterior.
Thus, one can expect the modification of the decay rate to be most
pronounced in those $\om$-regions where the polarizabilities of 
the second kind are large enough.
   
The relative magnitudes of the correction terms are governed not only 
by the transition energy but also are strongly  dependent on the position 
of the doped atom.
(This feature is absent in the radiative decay process.)
The parameter, which defines this dependence, is the ratio of the 
off-the-center atomic displacement $a$ from the fullerene center to the average 
radius $R$ of the cage. 
As $a/R$ increases the enhancement of the width occurs
for the transitions whose energies are in the vicinity of the fullerene 
surface plasmons energies of high multipolarity.  

We have demonstrated that the correction terms
as functions of both $\om$ and $a/R$ vary by orders of magnitude.
This can result in the dominance of the additional channels over the
direct Auger decay and can lead to pronounced broadening of the 
atomic emission lines.

On the basis of the developed formalism we carried out a case study 
for the system Sc$^{2+}$@C$_{80}^{6-}$.
It has been demonstrated that narrow resonances in the spectrum of an
isolated Sc$^{2+}$ in the energy range $\om = 30\dots 45$ eV
are noticeably broadened if the ion is located strongly off-the-center
($a/R\approx 0.5$).
Our model allowed us to carry out the quantitative analysis of the photoionization
of the endohedral complex Sc$_3$N@C$_{80}$.
We have demonstrated that due to the the non-central position of the three 
scandium atoms as well as to the fact that the multipolar surface plasmons energies
of the fullerene lie within the indicated range, the additional decay channels
can be responsible, at least partly, for the strong modification of the
photoionization spectrum profile detected experimentally.
The analysis of another reason of the lines broadening, which is also due
to the presence of the fullerene shell but different from those discussed
in this paper, will be published elsewhere.

\ack

We are grateful to Nikolai Cherepkov, Alfred M\"{u}ller and 
Stephan Schippers for the helpful discussions. 
This work was supported by the
European Commission within the Network of Excellence
project EXCELL (project no 515703).

%%%%%%%%%%%%%%%%%%%%%%%%%%%%%%%%%%%%%%%%%%%%%%%
\appendix

%%%%%%%%%%%%%%%%%%%%%%%%%%%%%%%%%%%%%%%%%%%%%%%%
\section{Multipole polarizabilities of hollow objects 
\label{MultipolPolar}}

In this Appendix we present general definitions and discuss basic 
properties of different types of multipole polarizabilities which
characterize the response of a 'hollow object' to an external electric 
field. 
By a hollow object we will understand any object which has at least
one cavity in its interior.
A fullerene $\rm C_N$, whose (average) radius $R$ greatly exceeds 
the radius of valence electrons in a carbon atom,
can be treated as a hollow object if one neglects 
the probability to observe electrons in most part of the fullerene interior.

%%%%%%%%%%%%%%%%%%%%%%%%%%%%%%%%%%%%%%%%%%%%%%%%
%\subsection{Static limit \label{MultipolPolar_1}}

%%%%%%%%%%%%%%%%%%%%%%%%%%%%%%%%%%%%%%%%%%%%%%%%
\subsection{Definitions
\label{MultipolPolar_1}}

Any system of electric charges becomes polarized if exposed to an external
electric field.
The polarization of this object leads to an additional electric field
whose potential (or/and strength) depend on the multipolarity of the
external field, on the distance from the system to the observation point, 
and on the (multipole) polarizability, i.e. the quantity which describes
the system's response to the external field.

If an object has no internal cavity (e.g., an atom, a simple molecule), 
it is natural to assume that a source of the external field 
(e.g., a point-like charge) is located outside the object.
Then one can distinguishes two different cases: 
the observation point is located (i) outside or (ii) inside the object.
The different quantities are responsible for the system's response
in these two cases.
Conventionally (see, e.g., \cite{Dalgarno1962}), the characteristic
associated with case (i) is called a polarizability, whilst 
case (ii) is described by the quantity termed "a shielding factor".
 
As discussed below, the response of a hollow object 
to the action of  an external electric field can be described in terms of 
{\it three} different quantities. 
In addition to the two mentioned ones, the third quantity corresponds to the
situation when both the external charge and the observation point are
located inside the object. 

Let us briefly discuss the quantum formalism which one can apply for 
a quantitative description of these quantities.
For the sake of simplicity, we consider the following model hollow object
(called below 'a fullerene'): a system of $N_e$ electrons located in the 
vicinity of the spherically-symmetric ionic cage of the radius $R$.
Let the fullerene be exposed to the field of a point charge $q$ located
in some point with the position vector $\bfr_q$. 

Expanding the operator of the Coulomb interaction 
$V = -\sum_{j=1}^{N_e}{q /|\bfr_j - \bfr_q|}$ between $q$ and the fullerene
electrons  in spherical harmonics,
one presents it in the form 
$V =-q\sum_{lm} \phi_{lm}(\bfr_q,\{\bfr_j\})$.
The multipole potentials $\phi_{lm}(\bfr_q,\{\bfr_j\})$
depends on whether the charge $q$ is located in the fullerene exterior
($r_q \gg \langle r_j \rangle$) or interior
($r_q \ll \langle r_j \rangle$):  

\begin{eqnarray}
\phi_{lm}(\bfr_q,\{\bfr_j\}) 
=
\sqrt{4\pi \over 2l+1}
\times
\cases{
{Y_{lm}^{*}(\bfr_q)\over r_q^{l+1}} \,
Q_{lm}(\{\bfr_j\})
& if  $r_q \gg \langle r_j \rangle $,
\\
r_q^l\,Y_{lm}^{*}(\bfr_q)\,
C_{lm}(\{\bfr_j\})
& if $r_q \ll \langle r_j \rangle $,
}
\label{MultipolPolar_1.02}
\end{eqnarray}
where
\begin{eqnarray}
\fl
Q_{lm}(\{\bfr_j\})
= 
\sqrt{4\pi\over 2l+1}\,
\sum_j r_j^{l}\, Y_{lm}(\bfr_j)\,,
\qquad
%\label{atom-cage.5}
%
C_{lm}(\{\bfr_j\})
= 
\sqrt{4\pi\over 2l+1}\,
\sum_j {Y_{lm}(\bfr_j) \over r_j^{l+1}}\,.
\label{MultipolPolar_1.03}
%\label{atom-cage.6}
\end{eqnarray}
The first one of these quantities, $Q_{lm}(\{\bfr_j\})$,
is expressed in terms of
so-called regular solid harmonics $ r^{l}\, Y_{lm}(\bfr)$
(see, e.g. \cite{BrinkSatchler,Jansen2000}), and is commonly associated
with the $2^l$-pole moment of a system of point unit charges 
(in quantum-mechanical terms, it is the operator of the $2^l$-pole moment).
This quantity allows one to calculate the multipole potential created by the 
system in its exterior.
The second quantity, $C_{lm}(\{\bfr_j\})$, 
written in terms of irregular solid harmonics $ r^{-l-1}\, Y_{lm}(\bfr)$ 
\cite{Jansen2000},
determines the multipole potential in the system interior
(see \cite{BrinkSatchler}).
To avoid ambiguity we call this quantity the '{\it interior} 
$2^l$-pole moment'.

Under the action of the multipolar field $-q\phi_{lm}(\bfr_q,\{\bfr_j\})$ 
the electron cloud in the fullerene becomes polarized.
This polarization causes the change in the potential, 
$\delta \phi_{lm}(\bfr)$, created by the electrons.  
Applying standard perturbation theory, one finds that $\delta \phi_{lm}(\bfr)$
is given by the following second-order matrix element:
\begin{eqnarray}
\delta \phi_{lm}(\bfr)
=
q
\sum_{n\neq0}
{
2 \left\langle 0 \Bigl| \phi_{lm}(\bfr_q,\{\bfr_j\}) \Bigr| n \right\rangle
\left\langle n \Bigl| \phi_{lm}^{*}(\bfr,\{\bfr_j\})\Bigr| n \right\rangle
\over 
\om_{n0}
}\,.
\label{MultipolPolar_1.04}
\end{eqnarray}
Here, $\phi_{lm}(\bfr,\{\bfr_j\})$ stands for the (operator of) 
multipole potential created by the electrons in the point $\bfr$.
Its explicit form follows from (\ref{MultipolPolar_1.02}) by means 
of the substitution $\bfr_q \to \bfr$.
The sum is carried out over the complete spectrum of the fullerene's 
excited states $n$ (the ground state $0$ is excluded) including the 
excitations into continuum, $\om_{n0}=\E_n-\E_0$ is the
excitation energy.  
  
Explicit form of the expression on
the right-hand side of (\ref{MultipolPolar_1.04}) 
depends not only on the location of the external charge $q$ 
with respect to $\langle r_j \rangle$, but also on the location of the 
observation point $\bfr$.
Considering different cases, one derives:
\begin{eqnarray}
\fl
\delta \phi_{lm}(\bfr)
=
q
{4\pi \over 2l+1}
Y_{lm}^{*}(\bfr_q)
Y_{lm}(\bfr)
\times
\cases{
{\alpha_l \over r_q^{l+1}r^{l+1} } 
& if  $r_q, r \gg \langle r_j \rangle $,\\
\beta_l\, { r^{l+1}\over r_q^{l+1} } 
& if  $r_q \gg \langle r_j \rangle \gg r$,\\
\walpha_l\, r_q^{l}r^{l}  
& if  $r_q, r \ll \langle r_j \rangle $.
}
\label{MultipolPolar_1.05}
\end{eqnarray}
The quantities $\alpha_l$, $\beta_l$ and $\walpha_l$ are given by
\begin{eqnarray}
\fl
\alpha_l 
=
\sum_{n\neq0}
{
2 
\Bigl|
\langle 0 \bigl| Q_{lm} \bigr| n \rangle
\Bigr|^2
\over 
\om_{n0}
}\,,
\quad
\beta_l 
=
\sum_{n\neq0}
{
2 
\langle 0 \bigl| Q_{lm} \bigr| n \rangle
\langle n \bigl| C_{lm}^{*} \bigr| 0 \rangle
\over 
\om_{n0}
}\,,
\quad
\walpha_l 
=
\sum_{n\neq0}
{
2 
\Bigl|
\langle 0 \bigl| C_{lm} \bigr| n \rangle
\Bigr|^2
\over 
\om_{n0}
} .
\label{MultipolPolar_1.06}
\end{eqnarray}
The first two of these are well-known in atomic physics (see, e.g., 
\cite{Dalgarno1962}).
The (static) {\it polarizability} $\alpha_l$ defines the change in the
field, created by a system, in the case when both the source charge and
the observation point are located in the system's exterior.
The quantity $\beta_l$ is called a {\it shielding factor}, and it 
defines the change in the field in the system's interior due to the 
charge re-distribution under the action of $q$ located in the exterior.
Atomic shielding factors appear when one analyzes the change of the field at
the nucleus due to the polarization of the outer shells electrons under
the action of external electric field. 

It is the third term, $\walpha_l$, which additionally appears 
when a hollow system is exposed to the external field. 
In this case, both the source charge $q$ and the observation point 
can be located in the cavity. 
To distinguish between $\alpha_l$ and $\walpha_l$ we call the latter
quantity the polarizability {\it of the second kind}.

The polarizability of the second kind naturally appears in the processes 
which involve the interaction of a subsystem {\it I} embedded by a hollow 
subsystem {\it II}.
An example would be an endohedral complex ${\rm A@C_N}$, where A stands 
for an atom.  
It can be shown that the dipole polarizability 
of the second kind $\walpha_1$ determines the van der Waals 
interaction between
the fullerene and the centrally placed atom 
\cite{Pyykko_EtAl2007,Pyykko_EtAl2010}. 
More general treatment of the van der Waals interaction between ${\rmC_N}$
and the atom, arbitrary placed in the fullerene interior, involves the 
polarizabilities $\walpha_l$ of higher multipolarities 
\cite{van_der_Waals}.
Dynamical screening of an endohedral atom
(see \cite{SolovyovConnerade2005,LoKorolSolovyov2007})
is another example of the process in which the polarizabilities 
$\walpha_l$ (more exactly, the 
{\it dynamic} polarizabilities $\walpha_l(\om)$, 
see (\ref{MultipolPolar_2.01})) manifest themselves.
In this process the presence of the fullerene dynamically screens
the confined atom from an external electromagnetic field, so that
the atom experiences a field that is either enhanced or suppressed depending
on the field frequency $\om$.
It can be shown (see  \cite{LoKorolSolovyov2009}) 
that to calculate the change in the atomic photoionization cross section
due to the dynamical screening one has to calculate the shielding factor of
the fullerene {\it and} the polarizabilities $\walpha_l(\om)$.
Equation (\ref{d_Auger_non_central5.13b}) 
in the main text demonstrates, that fullerene dynamic polarizabilities of the
second kind  determine the correction to the Auger decay rate of the
endohedral atom.

%%%%%%%%%%%%%%%%%%%%%%%%%%%%%%%%%%%%%%%%%%%%%%%%
\subsection{Dynamic multipole polarizabilities
\label{MultipolPolar_2}}

The formalism outlined above can be adjusted to the
case of the time-dependent external multipolar field.
To do this one carries out the Fourier transform of the external field and
then treats each Fourier component separately.
As a result, each of the quantities $\alpha_l$, $\walpha_l$ and $\beta_l$ 
acquires the dependence on $\om$ (the parameter of the Fourier transform or
the 'frequency' of the external field).
Explicit expressions  for the dynamic polarizabilities of all types are
as follows:
\begin{eqnarray}
\cases{
\alpha_l(\om)
=
\sum_{n\neq 0}
{2\om_{n0}\,\left|\langle 0|Q_{lm}| n \rangle \right|^2
\over \om_{n0}^2 - \om^2 -\i 0},
\\
\walpha_l(\om)
=
\sum_{n\neq 0}
{2\om_{n0}\,\left|\langle 0|C_{lm}| n \rangle \right|^2
\over \om_{n0}^2 - \om^2 -\i 0},
\\
\beta_l(\om)
=
\sum_{n\neq 0}
{2\om_{n0}\,
\langle 0|Q_{lm}| n \rangle
\langle n|C_{lm}^{*}| 0 \rangle
\over \om_{n0}^2 - \om^2 -\i 0}.
}
\label{MultipolPolar_2.01}
\end{eqnarray}
In the static limit $\om=0$ these formulae reduce to 
(\ref{MultipolPolar_1.06}).

Let us evaluate the asymptotic behaviour of
$\alpha_l(\om)$, $\walpha_l(\om)$ and $\beta_l(\om)$
in the region $\om \gg I_K$ ($I_K$ stands for the 
ionization potential of the most bound electrons in the system).
As a first step in deriving the asymptotic expressions one neglects 
$\om_{n0}^2$ in the denominators.
The resulting sums over the complete spectrum of the system
are carried out using standard methods of 
evaluation of the sum rules 
(see, e.g., \cite{Landau3}).
To calculate the gradients of regular $ r^{l}\, Y_{lm}(\bfr)$ and 
irregular $ r^{-l-1}\, Y_{lm}(\bfr)$ solid harmonic one can use
the formulae from Ch. 7.3.6 in \cite{VMX}.
The final result reads:
\begin{eqnarray}
\cases{
\lim_{\om\gg I_{K}}
\alpha^{(l)}(\om)
=
-{1\over \om^2}
\sum_{n}2\om_{n0}\,\left|\langle 0|Q_{lm}| n \rangle \right|^2
=
-
{N_e\over \om^2}\,
l\, \overline{r^{2l-2}} \,,
\\
\lim_{\om\gg I_{K}}
\walpha_l(\om)
=
-{1\over \om^2}
\sum_{n}2\om_{n0}\,\left|\langle 0|C_{lm}| n \rangle \right|^2
=
-{N_e \over \om^2}\,
(l+1)\, \overline{r^{-2l-4}}\,,
\\
\lim_{\om\gg I_{K}}
\beta_l(\om)
=
-{1\over \om^2}
\sum_{n}2\om_{n0}\,
\langle 0|Q_{lm}| n \rangle
\langle n|C_{lm}^{*}| 0 \rangle
=
0\,,
}
\label{MultipolPolar_2.02}
\end{eqnarray}
where $\overline{r^k}$ denotes the mean value
of $r^k$ in the ground state.

The first formulae from (\ref{MultipolPolar_2.02}) coincides with the
one obtained earlier in \cite{SolovyovConnerade2002}.
In the dipole case, $l=1$, it reproduces the well-known result, related
to the Thomas-Reiche-Kuhn dipole sum rule.

%%%%%%%%%%%%%%%%%%%%%%%%%%%%%%%%%%%%%%%%%%%%%%%%
\subsection{Multipole polarizabilities of a dielectric spherical shell 
\label{Sperical_layer}}

For reference, we present explicit expressions for the polarizabilities
$\alpha_l$, $\walpha_l$ and $\beta_l$ of a model hollow object: a spherical
shell made of a material with dielectric function $\E$.
The inner and outer radii of the shell are notated as $R_1$ and $R_2$, 
respectively.

To find the polarizability $\alpha_l$ and the shielding factor $\beta_l$ 
one places the external charge $q$ in the shell exterior ($r_q>R_2$) and
determines the additional potential $\delta \phi(\bfr)$ due to the 
polarization of the shell.
Multipolar components $\delta \phi_{lm}(\bfr)$ of this potential
are proportional either to $\alpha_l$ (if $r>R_2$) or to
$\beta_l$ (if $r<R_1$).
To find $\walpha_l$ one places the charge in the shell interior $r_q<R_1$
and looks for $\delta \phi_{lm}(\bfr)$ in the inner region $r<R_1$ .
As a result, the one derives
\begin{eqnarray}
\cases{
\alpha_l
=
l\, {(\E-1)(\E (l+1) +  l )
\over D_l}\, \Bigl(1 - \xi^{2l+1}\Bigr) R_2^{2l+1} ,
\\
\walpha_l
=
(l+1)\,
{(\E-1)(\E l + l+1) \over D_l}\,
{\Bigl(1- \xi^{2l+1}\Bigr) \over R_1^{2l+1}},
\\
\beta_l
=
-l (l+1)\,
{(\E-1)^2 \over D_l}\,
\Bigl(1-\xi^{2l+1}\Bigr)\,.
}
\label{Sperical_layer_1}
\end{eqnarray}
Here 
\begin{eqnarray}
\fl
D_l =
\Bigl(\E l + l+1 \Bigr)
\Bigl(\E (l+1) +  l \Bigr)
-l (l+1)(\E-1)^2
\xi^{2l+1}\, ,
\qquad
\xi = {R_1 \over R_2} < 1.
\label{Boundary2.6b}
\end{eqnarray}
The expression for the multipole polarizability $\alpha_l$, presented in
(\ref{Sperical_layer_1}), coincides with the formula obtained earlier in
\cite{LambinLucasVigneron1992}.

%%%%%%%%%%%%%%%%%%%%%%%%%%%%%%%%%%%%%%%%%%%%%%%%
\subsection{Multipole polarizabilities 
of a fullerene within plasmon resonance approximation  
\label{PRA}}

Photoionization experiments on C$_{60}$ and its positive ions reveal two 
giant resonances in their
photoionization spectra in the photon energy region $\approx 20\dots 40$ eV
\cite{BeckerEtAl_2004,ScullyEtAl_2004}.
These resonances are of a collective nature and are due 
to the excitations of plasmons by the photon field 
\cite{ScullyEtAl_2004,KorolSolovyov2007,BelyaevEtAl2009}.
The existence of two resonances can be explained \cite{KorolSolovyov2007}
in terms of
two coupled surface plasmons excited in a fullerene when it is considered
as a spherical layer of a finite thickness
\cite{LambinLucasVigneron1992,OstlinEtAl1993,AndersenBonderup2000,LoKorolSolovyov2007,LoKorolSolovyov2009,Lo_Thesis}.

In this section we apply this model to express each of the quantities
$\alpha_l$ and $\walpha_l$ from (\ref{Sperical_layer_1}) 
as a sum of two resonance terms corresponding to the two surface plasmons
modes. 
We call this approach as plasmon resonance approximation 
in accordance with the term used in the theory of 
plasmon resonances in metallic clusters 
(see, e.g., \cite{SolovyovConnerade2002,Solovyov2005} 
and references therein).

Let the fullerene valence electrons be distributed within a spherical shell
whose inner and outer radii are $R_1=R - \Delta R/2$ 
and $R_2=R + \Delta/2$, with $R$ standing for the (average) radius of 
the ionic core and $\Delta$ -- for the width of the electron cloud (see
figure \ref{atom-cage.fig}).
Within the framework of PRA one can model the dielectric function 
$\E\equiv \E(\om)$ of the fullerene electrons by a Lorentz-type 
dielectric function (see, e.g., \cite{AndersenBonderup2000}):
\begin{eqnarray}
\E  = 1 - {\om_p^2 \over \om^2 - \om_0^2 + \i \Gamma\om}\,,
\label{Polariz.1}
\end{eqnarray}
where $\Gamma$ is a damping constant, and the term $\om_0^2$ is included to
go beyond the free-electron gas approximation implied by the Drude model.
The plasma frequency $\om_p$ is found from $\om_p^2 = 4\pi N_e/V$, where
$V=4\pi(R_2^3-R_1^3)/3$ is the volume of the shell.

Using (\ref{Polariz.1}) in (\ref{Sperical_layer_1}), one represents
the dynamic polarizabilities $\alpha_l(\om)$ and $\walpha_l(\om)$ as follows: 
\begin{eqnarray}
\cases{
\alpha_{l}(\om)
=  
l \,
\calA_{l}(\om)\,
\overline{R^{2l-2}}\,,
\\
\walpha_{l}(\om)
=
(l+1) \,
\wcalA_{l}(\om)\,
\overline{R^{-2l-4}}\,,
}
\label{ModifiedPRA.4}
\end{eqnarray}
The factors $\overline{R^{2l-2}}$ and $\overline{R^{-2l-4}}$, which
stand for the mean $(2l-2)$th and $(-2l-4)$th powers of the radius of the 
electron cloud (cf., equation (\ref{MultipolPolar_2.02})), conveniently expose
the dependence of $\alpha_l$ and $\walpha_l$ on the shell radii:
\begin{eqnarray}
\fl
\overline{R^{2l-2}}
=
{3 \over 2l+1}\,
{1 - \xi^{2l+1} \over 1 - \xi^3}\,
R_2^{2l-2}\,,
\qquad
\overline{R^{-2l-4}}
=
{3 \over 2l+1}\,
{1\over R_2^3 R_1^{2l+1}}\,
{1 - \xi^{2l+1} \over 1 - \xi^3}\,.
\label{ModifiedPRA.6}
\end{eqnarray}

The $\om$-dependence of the polarizabilities is concentrated in the 
factors $\calA_{l}(\om)$ and $\wcalA_{l}(\om)$, which have similar 
resonance structure: 
\begin{eqnarray}
\fl
\qquad
\cases{
\calA_{l}(\om)
=
{N_{1l} \over  \om_{1l}^2 + \om_0^2 -\om^2 - \i \Gamma_{1l}\om}
+
{N_{2l} \over  \om_{2l}^2 + \om_0^2 -\om^2 - \i \Gamma_{2l}\om}
\\
\wcalA_{l}(\om)
=
{\wN_{1l} \over \om_{1l}^2 + \om_0^2  -\om^2 - \i \Gamma_{1l}\om}
+
{\wN_{2l} \over \om_{2l}^2 + \om_0^2  -\om^2 - \i \Gamma_{2l}\om}
}
\label{ModifiedPRA.5}
\end{eqnarray}
Here $\om_{1l}$ and $\om_{2l}$ are 
the surface-plasmon frequencies for a given multipolarity $l$. 
These pairs of eigenmodes
arise from the fullerene having a finite thickness and therefore
two surface charge densities. 
The eigenfrequency $\om_{1l}$ characterizes the symmetric mode of coupled
oscillations, 
in which the two surface charge densities oscillate in phase.
The eigenfrequency $\om_{2l}$ (which is larger than $\om_{1l}$) 
 corresponds to the antisymmetric mode, in which the oscillations 
are in antiphase
\cite{LambinLucasVigneron1992,LoKorolSolovyov2009}.
These frequencies are given by
\begin{eqnarray}
\om_{1l}^2
=
{\om_p^2\over 2(2l+1)} \left(2l+1 - p_l\right),
\qquad
\om_{2l}^2 
=
{\om_p^2\over 2(2l+1)} \left(2l+1 + p_l\right),
\label{Polariz.5}
\end{eqnarray}
where $p_l = \sqrt{1 + 4l(l+1)\xi^{2l+1}}$.
Note that for each multipolarity the frequencies $\om_{1l}$ and $\om_{2l}$
satisfy the relation $\om_{1l}^2 + \om_{2l}^2 = \om_p^2$, which 
is a consequence of a more general sum rule for the magnitudes of 
surface plasmon frequencies \cite{ApellEtAl1996}.

The quantities $N_{jl}$ and $\wN_{jl}$ ($j=1,2$) are the (multipolar) 
oscillator strengths associated with the modes $\om_{1l}$ and  $\om_{2l}$.
They satisfy the rule $N_{1l}+N_{2l} = \wN_{1l}+\wN_{2l} = N_e$, and, thus,
represent the  numbers of valence electrons participating in the
symmetric and antisymmetric oscillatory modes. 
These numbers are given by
 \begin{eqnarray}
N_{1l} = \wN_{2l} = N_e\,{p_l+1\over 2p_l},
\qquad
N_{2l} = \wN_{1l} = N_e \,{p_l-1\over 2p_l}\,.
\label{Polariz.3}
%\label{No_averaging.6c}
\end{eqnarray}
Comparing (\ref{ModifiedPRA.5}) and (\ref{Polariz.3}) one notices, that the
functions $\calA_{l}(\om)$ and $\wcalA_{l}(\om)$ differ only in
the number of oscillators in the two modes: the expression for 
$\wcalA_{l}(\om)$ can be obtained from that for $\calA_{l}(\om)$
by means of the substitution $N_{1l} \leftrightarrow N_{2l}$.

The quantities $\Gamma_{jl}$ ($j=1,2$) in (\ref{ModifiedPRA.5}) 
denote the half-widths of the surface plasmons multipole excitations. 
They determine the decay rate from collective excitation mode to the
incoherent sum of single-electron excitations. 
Within the framework of PRA they  (as well as the quantity $\om_0$)
are treated as parameters which can be either deduced from the experimental 
data or calculated separately 
\cite{Solovyov2005,GerchikovIpatovPolozkovSolovyov2000}. 

Within the PRA the shielding factor $\beta_l(\om)$ (see the last line
in (\ref{Sperical_layer_1})) acquires the form 
\cite{LoKorolSolovyov2009,Lo_Thesis}:
\begin{eqnarray}
\beta_{l}(\om)
=  
-{3l(l+1)\over 2l+1} \,
{1 - \xi^{2l+1} \over 1 - \xi^3}\,
{1\over R_2^3}\,
\calB_{l}(\om)\,.
\label{ShieldingFactor.1}
\end{eqnarray}
The factor $\calB_{l}(\om)$ has the following 
resonance structure: 
\begin{eqnarray}
\calB_{l}(\om)
=
{N_{1l}-N_{2l} \over  \om_{1l}^2 + \om_0^2 -\om^2 - \i \Gamma_{1l}\om}
-
{N_{1l}-N_{2l} \over  \om_{2l}^2 + \om_0^2 -\om^2 - \i \Gamma_{2l}\om}
\,.
\label{ShieldingFactor.2}
\end{eqnarray}
Note that, in contrast to $\calA_{l}(\om)$ and $\wcalA_{l}(\om)$,
the "oscillator strengths"  associated in $\calB_{l}(\om)$ with the
symmetric and asymmetric modes are the same and equal to  $N_{1l}-N_{2l}$.

%%%%%%%%%%%%%%%%%%%%%%%%%%%%%%%%%%%%%%%%%%%%%%%%%%%%%%%%%
\subsection{Dynamic multipole polarizabilities $\alpha_{l}(\om)$ 
and $\walpha_{l}(\om)$ of C$_{80}^{6-}$ 
\label{C80}} 

In this section we apply the plasmon resonance approximation 
to calculate dynamic polarizabilities $\alpha_{l}(\om)$ and 
$\walpha_{l}(\om)$ for the fullerene ion C$_{80}^{6-}$.

Taking into account the study of Nakao \etal \cite{NakaoKutitaFujita_1994} 
we assume spherical symmetry for the ionic cage and use the value 
$R=4.15$ \AA\, for the cage radius.
The thickness $\Delta R$ of the fullerene is set to 1.5 \AA, which was
obtained in \cite{RuedelEtAl2002} for C$_{60}$.
These values of $R$ and $\Delta R$, accompanied by the number of valence 
electrons $N_e=326$, lead to $\om_p = 37.0$ eV for the plasma frequency and
to $\xi = 0.69$ for the inner to outer radii ratio.

The value of $\om_0$ was chosen so as to equate the frequency 
of the  {\em symmetric dipole}  resonance (i.e., the resonance
frequency of the first term  in $A_1(\om)$, see (\ref{ModifiedPRA.5}))
to the experimentally measured position $\om_{\rm exp}$ of the plasmon
resonance peak in the photoionization spectrum.
Thus, the value $\om_0=13.9$ eV was calculated from  
$\sqrt{\om_{11}^2 + \om_0^2} = \om_{\rm exp}$, where 
$\om_{11}$ one calculates from (\ref{Polariz.5}), and 
we took  $\om_{\rm exp} = 21$ eV, 
which corresponds
to the experimental maximum of the photoionization cross section 
for C$_{80}^{+}$ 
\cite{MuellerEtAl2007,MuellerEtAl_Isacc2007}.

To calculate the half-widths the following rule was adopted:
$\Gamma_{jl} = \gamma_{j}\om_{jl}$ ($j=1,2$).  
For symmetric surface plasmon modes of all multipolarities 
the value of $\gamma_1$ was fixed at 0.16, which equals to the 
ratio of experimentally measured half-width $\Gamma_{\rm exp}$ 
to the position $\om_{\rm exp}$ of the dipole resonance   
\cite{MuellerEtAl2007,MuellerEtAl_Isacc2007}.
For the antisymmetric modes we used 
the value $\gamma_2=0.22$, which corresponds to the 
$\Gamma_{\rm exp}/\om_{\rm exp}$ ratio measured for the second plasmon
resonance in photoionization of C$_{60}^{q+}$ ions \cite{ScullyEtAl_2004}.

\begin{figure}[h]
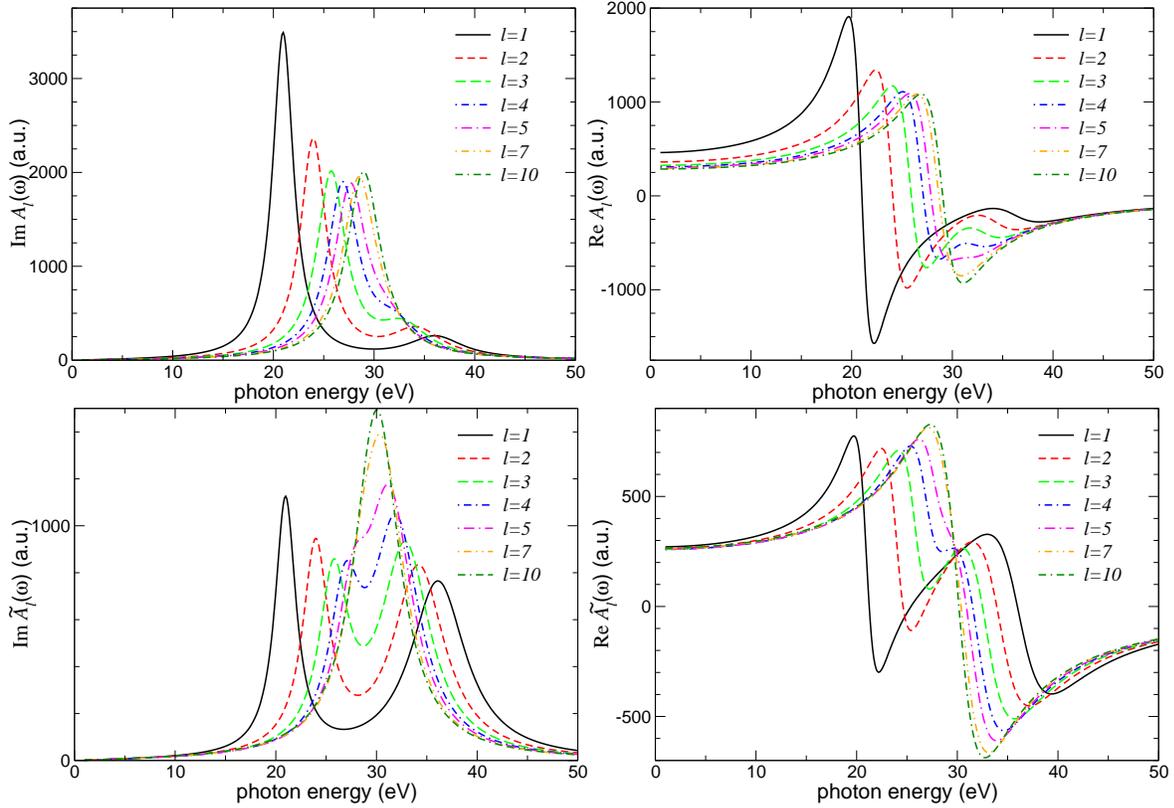

\includegraphics[clip,scale=0.31]{fig11a.eps}%{Im_A1_g2_020.eps}
\includegraphics[clip,scale=0.31]{fig11b.eps}%{Re_A1_g2_020.eps}
\\
\includegraphics[clip,scale=0.31]{fig11c.eps}%{Im_A2_g2_020.eps}
\includegraphics[clip,scale=0.31]{fig11d.eps}%{Re_A2_g2_020.eps}
\caption{
Imaginary and real parts of $\calA_l(\om)$ (upper graphs)
and $\wcalA_l(\om)$ (lower graphs) versus $\om$
calculated for  C$_{80}^{6-}$ and for several values of $l$ as 
indicated.
See explanation in the text for the parameters used in the calculations.
}
\label{A_L_C80.fig}
\end{figure}

The calculated dependences of real and imaginary parts of 
$\calA_l(\om)$ and $\wcalA_l(\om)$ on the photon energy $\om$
 for several values of $l$ are presented in figure \ref{A_L_C80.fig}.

%%%%%%%%%%%%%%%%%% Figures fig11a-d.eps
%                          Im \tilde{A}_l(\om) and Re \tilde{A}_l(\om)

We mention the following two features of the presented dependences.
Firstly, as it is seen from equation (\ref{Polariz.5}), with the 
growth of $l$ the frequencies $\om_{1l}$ and $\om_{2l}$
approach the same limit: $\om_{1l}$ goes to $\om_p/\sqrt{2}$ from 
below whereas $\om_{2l}$ -- from above.
As a result, the two resonances in both $\calA_l(\om)$ and $\wcalA_l(\om)$
merge as $l$ increases.
Secondly, the second (antisymmetric) surface plasmon
peak is quite pronounced for $\wcalA_l(\om)$ (i.e., in the
dynamic polarizability of the second kind) whereas in the case of
 $\calA_l(\om)$ its magnitude relative to the symmetric 
peak is much smaller. 
As $l$ increases (and until the peaks merge)
the  heights of symmetric and antisymmetric resonances nearly equalize 
in the case of 
$\wcalA_l(\om)$, whereas for  $\calA_l(\om)$ the antisymmetric peak 
quickly disappears.
The reason for such difference in the resonance curves behaviour is 
related to the different number of valence electrons participating 
in symmetric and antisymmetric modes, see (\ref{Polariz.3}).
The ratio $N_{2l}/N_{1l}$ rapidly decreases with $l$
(the ratio equals to $0.3$ for $l=1$ and to $0.04$ for $l=10$).
Hence, the number of oscillators associated with the antisymmetric mode in  
$\calA_l(\om)$,
which is $N_{2l}$, rapidly decreases, thus destroying the peak at $\om_{2l}$.
On the contrary, in $\wcalA_l(\om)$ where the antisymmetric mode is defined by
$N_{1l}$ oscillators, the relative intensity of its peak increases.

%%%%%%%%%%%%%%%%%%%%%%%%%%%%%%%%%%%%%%%%%%%%%%%%%%%%%%%%%
\subsection{Approximate relationship between $\alpha_l(\om)$  and
$\walpha_l(\om)$ for a fullerene
\label{Appendix_B}} 

To conclude Appendix let us mention a  transformation
which allows one to {\it approximately} relate the polarizabilities 
$\walpha_l(\om)$ and $\alpha_l(\om)$.
To simplify the algebra we consider the case of a spherically symmetric 
fullerene with the cage radius $R$ and the thickness $\Delta R$.

Assuming the strong inequality $\Delta R \ll R$ one reduces
the interior $2^l$-pole moment $Q_{lm}(\{\bfr_j\})$ to 
regular solid harmonics $ r^{l}\, Y_{lm}(\bfr)$
$Q_{lm}(\{\bfr_j\})$ (see equation (\ref{MultipolPolar_1.03})) as follows:
\begin{eqnarray}
\fl
C_{lm}(\{\bfr_j\})
= 
\sqrt{4\pi\over 2l+1}\,
\sum_j {r_j^{l} Y_{lm}(\bfr_j) \over r_j^{2l+1}}
\approx
{1\over R^{2l+1}}
\sqrt{4\pi\over 2l+1}\,
\sum_j r_j^{l} Y_{lm}(\bfr_j)
=
{Q_{lm}(\{\bfr_j\}) \over R^{2l+1}}\,.
\label{Appendix_B.01}
\end{eqnarray}
Using this result in (\ref{MultipolPolar_2.01})
one arrives at
\begin{eqnarray}
\walpha_l(\om)
\approx 
{\alpha_l(\om)  \over R^{4l+2}}\,,
\qquad
\beta_l(\om)
\approx
{\alpha_l(\om)\over R^{2l+1}}\,.
\label{Appendix_B.02}
\end{eqnarray}
Sometimes such approximation is used to estimate the 
dynamic response of a fullerene to the external field 
(see, e.g., (\cite{AmusiaBaltenkov2006})).
However, the accuracy of the approximation is subject to 
(comparatively) low $l$-values and
to a strong inequality $\Delta R \ll R$.
Thus, if one considers, for example, the fullerene C$_{60}$, 
for which $\Delta R/R \approx 0.4$, then 
equation (\ref{Appendix_B.02}) can be used only for the order-of-magnitude 
estimations and for $l\approx 1$.
The accuracy will increase for giant fullerenes (e.g., for C$_{960}$ 
where which $\Delta R/R \approx 0.1$).
Additionally, for low $l$-values this approximation neglects the 
difference between the profiles of symmetric and antisymmetric 
surface plasmon modes in $\walpha_l(\om)$, $\beta_l(\om)$
and $\alpha_l(\om)$
(to compare the profiles in $\alpha_l(\om)$ and in $\walpha_l(\om)$ -- 
see figure \ref{A_L_C80.fig}).

%%%%%%%%%% Acknowledgment
%%%%%%%%%%%%%%%%%%%%%%%%%%%%%%%%%%%%%%%%%%%%%%%%%%%%%%%%%%%%%%%%%%%%%%%%%%%%%%
\ack
This work was supported by the European Commission within the Network of
Excellence project EXCELL (project number 515703).

%%%%%%%%%% References
%%%%%%%%%%%%%%%%%%%%%%%%%%

%%%%%%%%%%%%%%%%%%%%%%%%%%%%%%%%%%%%%%%%%%%%%%%%%%%%%%%%%%%%%%%%%%%%%%%%
%%%%%%%%%%%%%%%%%%%%%%%%%%%%%%%%%%%%%%%%%%%%%%%%%%%%%%%%%%%%%%%%%%%%%%%%
%%%%%%%%%%%%%%%%%%%%%%%%%%%%%%%%%%%%%%%%%%%%%%%%%%%%%%%%%%%%%%%%%%%%%%%%
\end{document}